\documentclass[a4paper,UKenglish,cleveref, autoref, thm-restate]{lipics-v2021}

\pdfoutput=1 
\hideLIPIcs  

\title{CNOT-Optimal Clifford Synthesis as SAT}

\author{Irfansha Shaik}{Department of Computer Science, Aarhus University, Denmark\\Kvantify Aps, DK-2300 Copenhagen S, Denmark}{irfansha.shaik@cs.au.dk}{0000-0002-7404-348X}{}
\author{Jaco van de Pol}{Department of Computer Science, Aarhus University, Denmark}{jaco@cs.au.dk}{0000-0003-4305-0625}{}

\authorrunning{Irfansha Shaik and Jaco van de Pol}
  
\Copyright{Irfansha Shaik and Jaco van de Pol}

  

\supplement{}

\nolinenumbers 

\EventEditors{TODO}
\EventNoEds{2}
\EventLongTitle{28th International Conference on Theory and Applications of Satisfiability Testing (SAT 2025)}
\EventShortTitle{SAT 2025}
\EventAcronym{SAT}
\EventYear{2025}
\EventDate{August 12--15, 2025}
\EventLocation{Glasgow, Scotland}
\EventLogo{}
\SeriesVolume{305}
\ArticleNo{22}

\usepackage{amsmath}
\usepackage{amssymb}
\usepackage{blkarray}
\usepackage{booktabs}
\usepackage{mathtools}
\usepackage[qm]{qcircuit}
\usepackage{algorithm}
\usepackage[noend]{algorithmic}
\usepackage{graphicx}

\DeclareMathOperator{\EO}{EO}
\DeclareMathOperator{\AMO}{AMO}
\DeclareMathOperator{\ALO}{ALO}
\DeclareMathOperator{\ivar}{id}
\DeclareMathOperator{\hsvar}{hs}
\DeclareMathOperator{\shvar}{sh}
\DeclareMathOperator{\hvar}{h}
\DeclareMathOperator{\svar}{s}
\DeclareMathOperator{\hshvar}{hsh}
\DeclareMathOperator{\ctrlvar}{ctrl}
\DeclareMathOperator{\trgtvar}{trg}
\DeclareMathOperator{\cnotvar}{cnot}
\DeclareMathOperator{\xmatrixvar}{x}
\DeclareMathOperator{\zmatrixvar}{z}

\DeclareMathOperator{\last}{\textit{l}}
\DeclareMathOperator{\dxc}{dx}
\DeclareMathOperator{\dr}{dr}
\DeclareMathOperator{\dzc}{dz}
\DeclareMathOperator{\px}{px}
\DeclareMathOperator{\fx}{fx}
\DeclareMathOperator{\pz}{pz}

\DeclareMathOperator{\numqubits}{n}

\begin{document}

\maketitle

\begin{abstract}
Clifford circuit optimization is an important step in the quantum compilation pipeline.
Major compilers employ heuristic approaches. While they are fast, their results are often suboptimal.
Minimization of noisy gates, like 2-qubit CNOT gates, is crucial for practical computing.
Exact approaches have been proposed to fill the gap left by heuristic approaches.
Among these are SAT based approaches that optimize gate count or depth, but they suffer from scalability issues.
Further, they do not guarantee optimality on more important metrics like CNOT count or CNOT depth.
A recent work proposed an exhaustive search only on Clifford circuits in a certain normal form to guarantee CNOT count optimality.
But an exhaustive approach cannot scale beyond $6$ qubits.

In this paper, we incorporate search restricted to Clifford normal forms in a SAT encoding to guarantee CNOT count optimality.
By allowing parallel plans, we propose a second SAT encoding that optimizes CNOT depth.
By taking advantage of flexibility in SAT based approaches, we also handle
connectivity restrictions in hardware platforms, and allow for qubit relabeling.
We have implemented the above encodings and variations in our open source tool Q-Synth.

In experiments, our encodings significantly outperform existing SAT approaches on random Clifford circuits.
We consider practical VQE and Feynman benchmarks to compare with TKET and Qiskit compilers.
In all-to-all connectivity, we observe reductions up to 32.1\% in CNOT count and 48.1\% in CNOT depth.
Overall, we observe better results than TKET in the CNOT count and depth.
We also experiment with connectivity restrictions of major quantum platforms.
Compared to Qiskit, we observe up to 30.3\% CNOT count and 35.9\% CNOT depth further reduction.

\end{abstract}

\section{Introduction}
\label{sec:introduction}
Quantum Computing promises an alternative solution to some challenging
computational problems that are out-of-reach for classical computers.
While several competing quantum platforms exist in the current
Noisy Intermediate Scale Quantum (NISQ) era, they all come with different strengths and weaknesses.
Quantum programs must be compiled to low level quantum circuits
satisfying target hardware requirements before execution.
While most quantum platforms accept 1-qubit and 2-qubit gates, their native gate-sets can vary.
Currently, no quantum platform has the best qubit count, fidelity, and latency all together.
Circuit optimization can play a crucial role for practical quantum computing in both current NISQ
or future fault-tolerant processors.
For example,~\cite{ettenhuber2024calculatingenergyprofileenzymatic} observed that IonQ's
Aria quantum platform had a limit of 950 single qubit gates.
Only by using circuit optimization techniques they were able to run required circuits achieving
chemical accuracy.
For satisfying hardware requirements, Layout Synthesis and Circuit Synthesis are two main steps.
In Layout Synthesis, circuits are synthesized to handle hardware layout restrictions.
Often, not all qubits are connected in a quantum platform thus 2-qubit quantum gates can only be
scheduled on neighboring qubits.
Circuit Synthesis involves either synthesizing to required gate-set or optimizing some circuit metric
for practical quantum computing.

We consider circuit optimization in this paper.
Optimal synthesis of an arbitrary $\numqubits$-qubit circuit requires considering
$2^{\numqubits} \times 2^{\numqubits}$ unitary matrices of complex numbers.
While optimal circuit synthesis is ideal, it is a challenging computational problem~\cite{Nagarajan2021QuantumCircuitOptAO}.
Instead, peephole optimization is often used where \emph{easier} sub-circuits are optimized~\cite{DBLP:conf/ecai/ShaikP24}.
In this paper, we focus on an interesting subclass of circuits called \emph{Clifford circuits}.
Any circuit composed of $1$-qubit Hamard (H) and Phase (S) gates, and $2$-qubit Conditional-Not (CNOT) gates
is a Clifford circuit.
While polynomially simulatable, Clifford circuits capture important phenomena like entanglement and superposition,
and have applications in teleportation and dense quantum encoding~\cite{Aaronson_2004}.
Clifford circuits are also key in quantum error correction, required for the future fault-tolerant hardware.
Synthesis of Clifford circuits only requires $(2n) \times (2n+1)$ Boolean matrices, using the
\emph{stabilizer} formalism~\cite{Aaronson_2004}, instead of full unitary matrices.
Using peephole optimization, one can replace Clifford sub-circuits with their optimized counterparts.

There is no single optimization metric that can \emph{predict} actual hardware performance perfectly.
Quantitative metrics, like qubit fidelity, provide good predictions, but lead to numerical optimization problems.
Simple metrics like gate count, circuit depth are preferred for optimization~\cite{OLSQ2_2023, ShaikvdP2023, Sivarajah_2021, Wille_2023}.
In NISQ processors, 2-qubit CNOT gates are up to $10$ times more error-prone than 1-qubit gates.
Industrial compilers, like Qiskit~\cite{Qiskit} by IBM and TKET~\cite{Sivarajah_2021} by Quantinuum, apply Clifford optimization
reducing CNOT gate-count (cx-count) and CNOT depth (cx-depth).
Given a circuit with unary gates and CNOT gates, cx-depth is the maximum number of CNOT gates that are executed in series.
In a recent survey paper~\cite{nation2025benchmarkingperformancequantumcomputing}, the authors extensively compared major compilers on
cx-count/cx-depth reduction.
While several approaches exist for Clifford optimization, industrial compilers mainly focus
on heuristic approaches.
Exact optimization approaches unfortunately suffer from scalability problems.
Optimization of Clifford circuits is NP-hard~\cite{DBLP:conf/soda/JiangSTW0Z20}.
Even approximation of optimal synthesis is NP-hard~\cite{iwama2002transformation}, thus there is no efficient algorithm unless P = NP.
This undesirable gap between heuristic and exact approaches is well established in the literature~\cite{DBLP:conf/dac/WilleBZ19, Peham2023DepthOptimalSO}.

Thus, there is a need to improve the scalability of exact approaches.
In recent years, several SAT based approaches are proposed for problems like Layout Synthesis~\cite{DBLP:conf/dac/WilleBZ19, shaik2024optimal} and
CNOT synthesis~\cite{DBLP:conf/ecai/ShaikP24}.
In~\cite{Schneider2022ASE}, authors proposed a SAT encoding for cx-count optimization in Clifford circuits
based on bounded reachability.
Unfortunately, their encoding does not guarantee optimality due to the use of asymptotic
cx-count upper bound for termination criteria.
Despite being a near-optimal encoding, their approach still suffers from
scalability issues (even for $n>3$ qubits).
In~\cite{Peham2023DepthOptimalSO}, the authors proposed an improved SAT encoding instead optimizing circuit depth
that can handle up to $5$-qubit circuits.
While this improved scalability, their synthesis is focused on circuit depth, rather than cx-count and cx-depth.
Our experiments (\cref{sec:experiments}) show that synthesized circuits with minimal circuit depth can have worse cx-count/cx-depth,
even compared to heuristic tools.

Using some special properties of Clifford circuits, one can guarantee cx-count optimality.
In~\cite{Bravyi_2022}, authors proposed circuit synthesis restricted to certain normal forms.
They observed that by ignoring so-called phase updates one can guarantee cx-count optimality.
Using brute force search, they successfully generated a $2.1$TB database for all cx-count optimal
$6$-qubit circuits using up to $100$ days of compute.
While useful, generating such a database for beyond $6$-qubits is not practical~\cite{Bravyi_2022}.
Further, such an approach is not flexible i.e., a new database needs to be generated for each optimization metric/criteria.
For example, layout aware optimal Clifford circuits can vary depending on the platform layout restrictions.
Alternatively, flexibility of SAT like approaches can be used for efficient on-demand computation.

\paragraph*{Our Contribution}

In this paper, we incorporate the ideas proposed in~\cite{Bravyi_2022} to
obtain efficient SAT encodings that guarantee optimal Clifford circuits.
We provide the first \emph{cx-count} optimal SAT encoding based on bounded reachability.
Restricting the search to circuits in normal form reduces the search space (and consequently the makespan of the encoding) significantly.
Further, we propose the first \emph{cx-depth} optimal approach to Clifford circuit synthesis, by adapting the SAT encoding.
We have extended our open source tool Q-Synth\footnote{Q-Synth v5 tool with source code, benchmarks,
and scripts~\url{https://github.com/irfansha/Q-Synth}} with version 5 implementing the above encodings.
For an experimental comparison, we propose three experiments.
In Experiment 1, we compare existing SAT encodings with ours on random Clifford circuits of $3$ to $7$ qubits.
For both cx-count/cx-depth optimization, we significantly outperform previous approach while solving optimally.
We show that our approach can optimally solve $4$ out of $5$ $7$-qubit circuits for the first time for cx-depth optimization.
In Experiment 2, we demonstrate our effectiveness on practical VQE and Feynman~\cite{feynman2016} benchmarks via peephole optimization.
We generate 12 VQE benchmarks of 8 and 16 qubits, and 28 T-gate optimized Feynman benchmarks of $5$ to $24$ qubits.
Given a 10-minute time limit, we consistently outperform TKET, and 
we observe the best results when TKET+Q-Synth are used together.
In Experiment 3, we focus on layout aware re-synthesis of practical VQE and Feynman benchmarks.
We first use Qiskit~\cite{Qiskit} to map VQE and Feynman benchmarks onto 54-qubit Sycamore~\cite{arute2019quantum}, 80-qubit Rigetti~\cite{computingrigetti}, and 127-qubit Eagle~\cite{chow2021ibm} platforms.
We then re-synthesize each benchmark with Q-Synth giving a 10-minute time limit.
Overall, we observe significant reduction on all platforms in both cx-count and cx-depth optimizations.
In VQE benchmarks, we observe a reduction of up to 19.3\% cx-count and 27.4\% cx-depth.
In Feynman benchmarks, we observe a reduction of up to  30.3\% cx-count and 35.9\% cx-depth.
Our experiments indicate that there is a place for SAT like approaches in the quantum compilation pipeline.

\section{Preliminaries}
\label{sec:preliminaries}

\subsection{Clifford Circuits}
\label{subsec:cliffordcircuits}

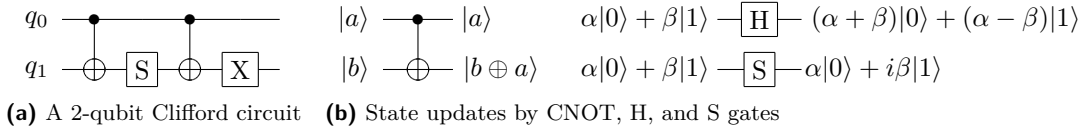
\begin{figure}[thbp]
\centering
\begin{subfigure}{0.3\textwidth}
\centering
\scalebox{1}{
\Qcircuit @C=0.8em @R=0.6em @!R { \\
\lstick{{q}_{0}} & \ctrl{1} & \qw & \ctrl{1} & \qw & \qw\\
\lstick{{q}_{1}} & \targ    & \gate{\mathrm{S}} & \targ & \gate{\mathrm{X}} & \qw
}}
\caption{A 2-qubit Clifford circuit}
\label{fig:orgcircuit}
\end{subfigure}%
\begin{subfigure}{0.7\textwidth}
\scalebox{1}{
\Qcircuit @C=0.9em @R=0.6em @!R { \\
& &\lstick{|a\rangle} & \ctrl{1} & \qw & |a\rangle           & & & & & & & \alpha |0\rangle + \beta |1\rangle
& & & & \gate{\mathrm{H}}  & \qw &&&&&& (\alpha+\beta) |0\rangle + (\alpha-\beta) |1\rangle\\
& &\lstick{|b\rangle} & \targ    & \qw & &|b \oplus a\rangle & & & & &   &  \alpha |0\rangle + \beta |1\rangle
& & & & \gate{\mathrm{S}} & \qw & & & \alpha |0\rangle + i\beta |1\rangle
}}
\caption{State updates by CNOT, H, and S gates}
\label{fig:stateupdates}
\end{subfigure}
\caption{Example Clifford circuit, and an illustration of state updates by Clifford gates.}
\label{fig:example}
\end{figure}

In classical computing, the fundamental unit of information, a classical bit,
has only two states $0$ and $1$.
In Quantum computing, a quantum bit instead is a superposition of $0$ and $1$.
One can represent a qubit state as a vector $\alpha |0\rangle + \beta |1\rangle $ where
$\alpha, \beta \in \mathbb{C}$ and $|\alpha|^2 + |\beta|^2 = 1$.
States over $\numqubits$-qubits live in the tensor product space on all vectors on $\numqubits$ qubits.
Quantum gates change the state by acting on one or more qubits.
Due to practical difficulties of implementing muti-qubit quantum gates, quantum platforms typically only apply 1-qubit and 2-qubit gates.
High level quantum programs are decomposed to low level circuits with 1- and 2-qubit gates before execution.
In this paper, we focus on Clifford circuits made of \{CNOT, H, S\} gates.
Figure~\ref{fig:stateupdates} shows how the Clifford gates change the state of qubits.
A CNOT gate entangles two qubits by applying an eXclusive OR. The H gate brings a qubit in a superposition,
and the S gate applies a phase change to a qubit.
Pauli gates \{X, Y, Z\} are composed of Clifford gates, where X = HSSH, Z = SS and Y = XZ.
We will use Pauli gates later in the paper for so-called relative phase recovery.
Figure~\ref{fig:orgcircuit} shows an example Clifford circuit with \{CNOT, S, X\} gates.
We refer interested readers to~\cite{Nielsen_Chuang_2010} for further understanding of quantum gates and quantum states.

Interestingly, Clifford gates are not universal and can be simulated in polynomial time and space.
In~\cite{Aaronson_2004}, Aaronson and Gottesman proposed \emph{stabilizer} formalism to efficiently represent Clifford circuits.
For an $\numqubits$-qubit Clifford circuit, we only need a $2\numqubits \times (2\numqubits+1)$ Boolean matrix, called \emph{tableau}.
Two Clifford circuits are equivalent (up to global phase) if their corresponding tableaux are equal~\cite{Bravyi_2022}.
The following matrix shows the structure of a $\numqubits$-qubit tableau:
\[
\setlength{\arraycolsep}{2pt}
\renewcommand{\arraystretch}{1}
\begin{blockarray}{ccccccc}
  q_0 & \dots & q_{n-1} & q_0 & \dots & q_{n-1} \\
  \begin{block}{(ccc|ccc|c)}
  x_{00} & \dots  & x_{0(n-1)} & z_{00} & \dots  & z_{0(n-1)} & r_{0}\\
  \vdots & \ddots & \vdots & \vdots & \ddots & \vdots & \vdots\\
  x_{(2n-1)0} & \dots  & x_{(2n-1)(n-1)} & z_{(2n-1)0} & \dots  & z_{(2n-1)(n-1)} & r_{2n-1}\\
  \end{block}
  \end{blockarray}  
\]
An $\numqubits$-qubit tableau is made of $x_{2n \times n}$, $z_{2n\times n}$, and $r_{2n \times 1}$ Boolean matrices.
A tableau essentially represents so-called \emph{stabilizer} and \emph{destabilizer} generators.
We refer to~\cite{Aaronson_2004} for a detailed explanation on stabilizer generators, Clifford circuits, and relevant proofs.
For the scope of this paper, it is sufficient to understand how a tableau can be computed given a Clifford circuit.
The initial tableau, representing an empty circuit, is defined as $x_{ii} = z_{(n+i)i} = 1$ (for $0\leq i<n$)
and every other cell is $0$.
\begin{table}[tbhp]
  \caption{Tableau update rules for every row $i$,
  when applying Clifford gates to qubits $a$ and $b$}
  \label{tb:updaterules}
  \centering
  \begin{tabular}{lcccccc}
  \toprule
              & \multicolumn{3}{c}{Base Gates} & \multicolumn{3}{c}{Pauli gates}\\
              \cmidrule(lr){2-4} \cmidrule(lr){5-7}
          & $\text{CNOT}_{a,b}$        & $\text{H}_{a}$  & $\text{S}_{a}$ & $\text{X}_{a}$ & $\text{Y}_{a}$ & $\text{Z}_{a}$\\
  $x_{ia}$ & $x_{ia}$                & $z_{ia}$ & $x_{ia}$ & $x_{ia}$ & $x_{ia}$ & $x_{ia}$\\
  $z_{ia}$ & $z_{ia} \oplus z_{ib}$ & $x_{ia}$ & $x_{ia} \oplus z_{ia}$ & $z_{ia}$& $z_{ia}$ & $z_{ia}$\\
  $x_{ib}$ & $x_{ia} \oplus x_{ib}$ & - & - & -& - & -\\
  $z_{ib}$ & $z_{ib}$ & - & - & -& - & -\\
  $r_{i}$ & $ r_{i} \oplus x_{ia} z_{ib} (x_{ib} \oplus z_{ia} \oplus 1)$ & $r_{i} \oplus x_{ia}z_{ia}$
  & $r_{i} \oplus x_{ia}z_{ia}$ & $r_{i} \oplus z_{ia}$ & $r_{i} \oplus x_{ia}\oplus x_{ib}$ & $r_{i} \oplus x_{ia}$\\
  \bottomrule
  \end{tabular}
  \end{table}

For each gate in a given circuit, we update every row of the tableau according to the rules in \cref{tb:updaterules}; this corresponds to column additions modulo 2.
For our example Clifford circuit from \cref{fig:orgcircuit}, the following sequence of matrices represent tableau updates:
\[
\setlength{\arraycolsep}{1.5pt}
\renewcommand{\arraystretch}{0.7}
\left(
\begin{array}{cc|cc|c}
1 & 0 & 0 & 0 & 0\\
0 & 1 & 0 & 0 & 0\\
0 & 0 & 1 & 0 & 0\\
0 & 0 & 0 & 1 & 0
\end{array}
\right)
\xrightarrow[q_0,q_1]{\text{CX}}
\left(
\begin{array}{cc|cc|c}
1 & \mathbf{1} & 0 & 0 & 0\\
0 & 1 & 0 & 0 & 0\\
0 & 0 & 1 & 0 & 0\\
0 & 0 & \mathbf{1} & 1 & 0
\end{array}
\right)
\xrightarrow[q_1]{\text{S}}
\left(
\begin{array}{cc|cc|c}
1 & 1 & 0 & \mathbf{1} & 0\\
0 & 1 & 0 & \mathbf{1} & 0\\
0 & 0 & 1 & 0 & 0\\
0 & 0 & 1 & 1 & 0
\end{array}
\right)
\xrightarrow[q_0,q_1]{\text{CX}}
\left(
\begin{array}{cc|cc|c}
1 & \mathbf{0} & \mathbf{1} & 1 & 0\\
0 & 1 & \mathbf{1} & 1 & 0\\
0 & 0 & 1 & 0 & 0\\
0 & 0 & \mathbf{0} & 1 & 0
\end{array}
\right)
\xrightarrow[q_1]{\text{X}}
\left(
\begin{array}{cc|cc|c}
1 & 0 & 1 & 1 & \mathbf{1}\\
0 & 1 & 1 & 1 & \mathbf{1}\\
0 & 0 & 1 & 0 & 0\\
0 & 0 & 0 & 1 & \mathbf{1}
\end{array}
\right)
\]
The leftmost matrix corresponds to the initial tableau and the rightmost matrix is our target tableau.
Optimal synthesis of Clifford circuits can now be transformed to a graph search problem.
Each node of the graph is labelled with a tableau, and edges labeled with Clifford gates correspond to the update rules.
Synthesizing the optimal number of gates boils down to finding the shortest path in this graph.
On the other hand, optimizing the number of CNOT gates is not so straightforward. We will
revisit this in \cref{sec:revisitingcliffordnormalforms}.

\subsection{Layout Restrictions and Qubit Relabeling}
\label{subsec:layoutrestrictions}

Most existing Clifford synthesis approaches~\cite{Bravyi_2022,Sivarajah_2021,Qiskit,Peham2023DepthOptimalSO}
assume all-to-all qubit connectivity.
Under such an assumption, local rewrite rules can be used effectively.
While useful, such structure breaks down in case of restricted qubit connectivity.
Thus, compilers like Qiskit and TKET apply Clifford synthesis before the layout synthesis phase, leading to suboptimal results overall.
Some heuristic approaches~\cite{DBLP:journals/corr/abs-2205-00724} exist that handle connectivity restrictions in related problems like CNOT synthesis.
In~\cite{DBLP:conf/ecai/ShaikP24}, we showed that connectivity restrictions can be encoded elegantly using SAT in CNOT synthesis.
Given a platform connectivity graph, we will only allow 2-qubit gates on neighboring qubits.
Thus, our approach can be used post layout synthesis, allowing further reduction.

In tableau representation, notice that the columns are labelled with qubits.
Allowing permutation of columns essentially corresponds to relabeling qubits.
In a relaxed notion of equivalence, two Clifford circuits are equivalent if their tableau are equivalent up to a column permutation.
Allowing any column permutation of initial tableau can result in better circuits.
Major compilers like TKET also allow qubit relabeling in the context of all-to-all connectivity.
In~\cite{DBLP:conf/ecai/ShaikP24}, we showed that permutation of a similar matrix can be elegantly encode using Exactly-One constraints.
We adapt the same idea for Clifford synthesis, we encode permutation of columns in initial tableau using cardinality constraints.

\section{Revisiting Clifford Normal Forms for Optimal CNOT Synthesis}
\label{sec:revisitingcliffordnormalforms}

In earlier sections, we briefly discussed that a simple bounded reachability encoding
is sufficient for gate count optimality.
For cx-count optimization, existing SAT approach in~\cite{Schneider2022ASE} proposed a MAXSAT like formulation
to optimize cx-count for the given gate count.
While they can synthesize circuits with better cx-count, they do not ensure cx-count optimality.
Consider our example circuit~\ref{fig:orgcircuit} with $4$ gates and $2$ CNOTs.
If we run QMAP tool which implements SAT encoding in~\cite{Schneider2022ASE} on our circuit,
it fails to produce a circuit with better cx-count.
This makes sense as there does not exist a $4$ gate circuit with better cx-count.
However, if we consider circuits with higher gate count, there does exist a circuit
with a single CNOT gate as shown in Figure~\ref{fig:1cnotnormalformphase}.
So the given number of Clifford gates in the input circuit is not a valid upperbound. Also, the upperbound of $\Theta(n^2/log(n))$ gates for an
$n$-qubit Clifford circuit \cite{patel2003efficientsynthesislinearreversible,Aaronson_2004} is only asymptotically sharp, so it doesn't provide a reliable upper bound for concrete $n$.

To ensure cx-count optimality, we need to reformulate the search problem.
Let $q,q'$ be two qubits, we define the set of \emph{entangling} sequences
$E = \{\Psi\text{CNOT}_{q,q'}\mid \Psi \in \{\text{H}_q, \text{H}_{q'}, \text{S}_q, \text{S}_{q'}\}^*\}$.
Each entangling sequence, by definition, adds a single CNOT gate to the circuit.
A $d$-CNOT optimal circuit implies that there does not exist a Clifford circuit with $<d$ entangling sequences.
This modification results in a graph search problem where nodes are still labelled with tableaux.
Edges on the other hand are labelled with entangling sequences instead of individual gates.
Finding an optimal CNOT circuit corresponds to finding the shortest path in the new graph.
While this new formulation guarantees CNOT optimality, the problem of arbitrary single gate sequences still remains.
Taking advantage of equivalences between Clifford circuits,
\cite{Bravyi_2022} proposed a normal form that considers only $9$ unique entangling sequences.
In the following section, we will revisit observations made in~\cite{Bravyi_2022} and adapt in the context of a SAT encoding.

\subsection{Clifford Normal Forms}
\label{subsec:cliffordnormalforms}

Remember that a tableau is made of $x,z,r$ matrices.
\cite{Bravyi_2022} observed that two tableau with same $x$ and $z$ matrices have same optimal cx-count.
From Table~\ref{tb:updaterules}, we can see that Pauli gates only update $r$ column.
Indeed, one can synthesize any circuit with differing $r$ column by appending Pauli gates at the front.
The authors in~\cite{Bravyi_2022} consider all circuits with differing $r$ column as an equivalence class.
For now, we ignore the $r$ column updates aka relative phase updates.
This makes the tableau update rules simpler, resulting in fewer unique $1$-qubit gate sequences.
Turns out there are not many sequences of H and S gates that result in unique tableau state.
Clearly, HH and SS cancel out thus only alternating H and S sequences of gates are interesting.
So the interesting sequences are \{I, H, S, HS, SH, HSH, SHS, HSHS, SHSH, ...\}.
Further, HSH and SHS result in the same tableau state.
Using the equivalence of HSH and SHS, we can simplify any non-trivial 1-q gate sequence to a sequence of length less than 4.
For example, consider HSHS gate sequence and replace HSH with SHS resulting in SHSS.
Since SS cancel out, we are left with SH sequence.
Any such sequence can be replaced with one of the 6 \{I, H, S, HS, SH, HSH\} sequences, defined as 1-q unique sequences.
Considering these $6$ sequences on each qubit followed by a CNOT results in 36 sequences.
\cite{Bravyi_2022} showed that only the following $9$ sequences are unique entangling sequences.
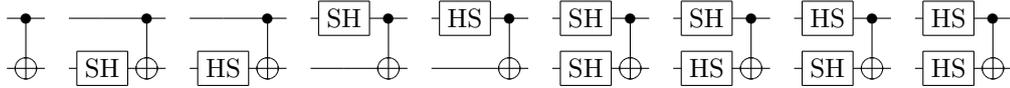
\[
\Qcircuit @C=0.3em @R=0.6em @!R {
& \ctrl{1}           & \qw &&&
&\qw                 & \ctrl{1} & \qw &&&
&\qw                 & \ctrl{1} & \qw &&&
&\gate{\mathrm{SH}}  & \ctrl{1} & \qw &&&
& \gate{\mathrm{HS}} & \ctrl{1} & \qw &&&
& \gate{\mathrm{SH}} & \ctrl{1} & \qw &&&
&\gate{\mathrm{SH}}  & \ctrl{1} & \qw &&&
& \gate{\mathrm{HS}} & \ctrl{1} & \qw &&&
&\gate{\mathrm{HS}}  & \ctrl{1} & \qw\\
& \targ              & \qw &&&
& \gate{\mathrm{SH}} & \targ & \qw &&&
& \gate{\mathrm{HS}} & \targ & \qw &&&
& \qw                & \targ & \qw &&&
& \qw                & \targ & \qw &&&
&\gate{\mathrm{SH}}  & \targ & \qw &&&
&\gate{\mathrm{HS}}  & \targ & \qw &&&
&\gate{\mathrm{SH}}  & \targ & \qw &&&
&\gate{\mathrm{HS}}  & \targ & \qw
}
\]
They showed that each of the $36$ entangling sequences can be rewritten as one of the above $9$ sequences
followed by one of the 1-q unique sequences on each qubit.
Note that, a CNOT gate $\text{CNOT}_{j,i}$ can be rewritten as $\text{H}_i\text{H}_j\text{CNOT}_{i,j}\text{H}_i\text{H}_j$.
Thus, we can rewrite all CNOT gates in a given circuit such that the control qubit is less than its target qubit.
Now, let us suppose, we have an entangling sequence in the form $\text{A}_{i}\text{B}_{j}\text{CNOT}_{i,j}$
where A, B are elements of 1-q unique sequences.
Such a sequence can be rewritten to the form $\text{A'}_{i}\text{B'}_{j}\text{CNOT}_{i,j}\text{C}_{i}\text{D}_{j}$
where A', B' are in \{I, HS, SH\} and C, D are 1-q unique sequences.
Using these rewrite rules, from start to end of the circuit, one can push 1-q sequences that are not
\{I, HS, SH\} to the last layer.
The final rewritten circuit will only have the above 9 unique entangling sequences followed by 1-q sequences on every qubit.

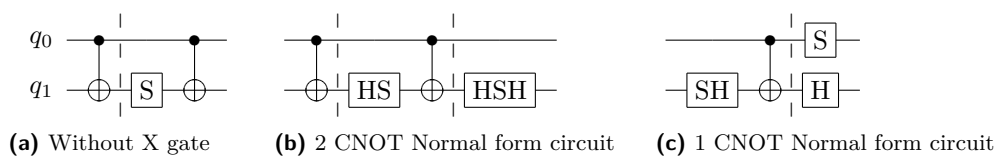
\begin{figure}[!b]
  \centering
  \begin{subfigure}{0.25\textwidth}
  \centering
  \scalebox{1}{
  \Qcircuit @C=0.8em @R=0.6em @!R { \\
  \lstick{{q}_{0}} & \ctrl{1}   \ar@{--}[]+<0.8em,1em>;[d]+<0.8em,-1em> & \qw & \ctrl{1} & \qw\\
  \lstick{{q}_{1}} & \targ    & \gate{\mathrm{S}} & \targ & \qw
  }}
  \caption{Without X gate}
  \label{fig:phaseignoredexample}
  \end{subfigure}%
  \begin{subfigure}{0.35\textwidth}
  \scalebox{1}{
  \Qcircuit @C=0.8em @R=0.6em @!R { \\
   & \ctrl{1}\ar@{--}[]+<0.8em,1em>;[d]+<0.8em,-1em> & \qw & \ctrl{1} \ar@{--}[]+<0.8em,1em>;[d]+<0.8em,-1em>& \qw & \qw\\
   & \targ    & \gate{\mathrm{HS}} & \targ & \gate{\mathrm{HSH}} & \qw
  }}
  \caption{2 CNOT Normal form circuit}
  \label{fig:2cnotnormalform}
  \end{subfigure}
  \begin{subfigure}{0.35\textwidth}
  \scalebox{1}{
  \Qcircuit @C=0.8em @R=0.6em @!R { \\
   & \qw & \ctrl{1} \ar@{--}[]+<0.8em,1em>;[d]+<0.8em,-1em>& \gate{\mathrm{S}} & \qw\\
   & \gate{\mathrm{SH}} & \targ & \gate{\mathrm{H}} & \qw
  }}
  \caption{1 CNOT Normal form circuit}
  \label{fig:1cnotnormalform}
  \end{subfigure}  
  \caption{Equivalent Clifford circuits of \cref{fig:orgcircuit}, ignoring relative phase}
  \label{fig:eqexample}
  \end{figure}

Consider our example circuit without X gate as in~\ref{fig:phaseignoredexample}.
First CNOT is part of unique entangling sequence, however, S gate followed by second CNOT is not.
One can replace $\text{S}_{q_1}\text{CNOT}_{q_0,q_1}$ by $\text{HS}_{q_1}\text{CNOT}_{q_0,q_1}\text{HSH}_{q_1}$.
The resulting circuit as shown in~\ref{fig:2cnotnormalform} has both CNOTs within unique sequences
followed by 1-q unique sequences.
Figure~\ref{fig:1cnotnormalform} shows an equivalent circuit with only 1 CNOT.
We illustrate entangling sequences in Figure~\ref{fig:eqexample} with dotted lines.
Authors in~\cite{Bravyi_2022} consider all circuit with same entangling sequences as part of an equivalence class.
However, for a SAT formulation we need to encode the final layer of 1-q sequences explicitly.

Using the above observations, synthesizing $d$-CNOT circuit corresponds to
finding $d$ entangling sequences followed by 1-q unique sequences.
For synthesizing $d$-CNOT optimality, we need to show there does not exist a normal form circuit
with $<d$ entangling sequences.
We can easily adapt these observations to handle CNOT-depth optimal synthesis.
Allowing parallel entangling sequences on independent qubits at each step allows CNOT-depth optimization.
This notion is exactly same as $\forall$-step parallel plans~\cite{DBLP:conf/kr/KautzMS96} in the context of SAT based planning.
In other words, sequential encoding corresponds to cx-count optimization whereas parallel encoding
corresponds to cx-depth optimization.
In Section~\ref{sec:satencodings}, we present encodings for both cx-count and cx-depth optimization.

\paragraph*{Phase Recovery for Peephole Setting}
Earlier in this Section, we ignored phase updates when synthesizing Clifford gates.
Thus, the optimized Clifford circuit (either by cx-count or cx-depth) will have a different
$r$-column in the tableau matrix.
While relative phase can be ignored for pure Clifford circuit optimization, 
for general quantum circuits (i.e., in particular for peephole-optimization),
we need to reconstruct the relative phase ($r$-column).
This can be achieved separately for each qubit, by appending single Pauli gates to the optimized circuit.
Appending Pauli gates at the beginning of the circuit is same as so-called \emph{Pauli left multiplication}.
In plain words, adding either X or Z Pauli gates at the beginning of the circuit flips the target tableau $r$-column.
Remember that Pauli gates do not change $x, z$ matrices but only act on the $r$-column.
Phase recovery is then figuring out which Pauli gates to be applied to recover our target $r$-column.
We observe that applying a $\text{Z}_i$ gate flips $r_i$ whereas $\text{X}_i$ gate flips $r_{n+i}$ in the target tableau.

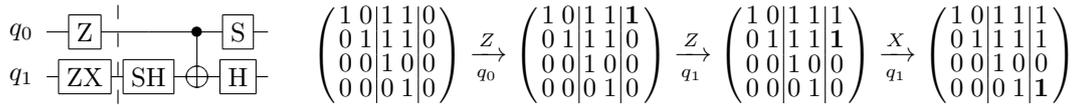
\begin{figure}[b!]
\centering
\begin{subfigure}{0.27\textwidth}
\centering
\scalebox{1}{
\Qcircuit @C=0.45em @R=0.4em @!R {
\lstick{{q}_{0}} & \gate{\mathrm{Z}} \ar@{--}[]+<1.25em,1em>;[d]+<1.25em,-1em>  & \qw                & \ctrl{1} & \gate{\mathrm{S}} & \qw\\
\lstick{{q}_{1}} &\gate{\mathrm{ZX}} & \gate{\mathrm{SH}} & \targ    & \gate{\mathrm{H}} & \qw
}}
\vspace{1em}
\caption{Optimal circuit}
\label{fig:1cnotnormalformphase}
\end{subfigure}
\begin{subfigure}{0.72\textwidth}
\vspace{-1em}
\[
\setlength{\arraycolsep}{1.3pt}
\renewcommand{\arraystretch}{0.7}
\left(
\begin{array}{cc|cc|c}
1 & 0 & 1 & 1 & 0\\
0 & 1 & 1 & 1 & 0\\
0 & 0 & 1 & 0 & 0\\
0 & 0 & 0 & 1 & 0
\end{array}
\right)
\xrightarrow[q_0]{Z}
\left(
\begin{array}{cc|cc|c}
1 & 0 & 1 & 1 & \mathbf{1}\\
0 & 1 & 1 & 1 & 0\\
0 & 0 & 1 & 0 & 0\\
0 & 0 & 0 & 1 & 0
\end{array}
\right)
\xrightarrow[q_1]{Z}
\left(
\begin{array}{cc|cc|c}
1 & 0 & 1 & 1 & 1\\
0 & 1 & 1 & 1 & \mathbf{1}\\
0 & 0 & 1 & 0 & 0\\
0 & 0 & 0 & 1 & 0
\end{array}
\right)
\xrightarrow[q_1]{X}
\left(
\begin{array}{cc|cc|c}
1 & 0 & 1 & 1 & 1\\
0 & 1 & 1 & 1 & 1\\
0 & 0 & 1 & 0 & 0\\
0 & 0 & 0 & 1 & \mathbf{1}
\end{array}
\right)
\vspace{-2em}
\]
\caption{Tableau transforms to correct phase}
\label{fig:phasetransformationtableau}
\end{subfigure}
\caption{Phase recovery applied to the Clifford circuit of \cref{fig:1cnotnormalform}}
\label{fig:phaseexample}
\end{figure}

Let $r,r'$ be $r$-columns of optimized and target tableau of an $\numqubits$-qubit circuit.
Here, we present a simple algorithm that computes the sequence of X and Z gates that recovers the relative phase.
\cref{alg:recovery} clearly runs in linear time in the number of qubits.
Consider the CNOT optimal example circuit in Figure~\ref{fig:1cnotnormalform}: the tableau deviates in the $r$-column.
Appending X and Z gates as shown in Figure~\ref{fig:1cnotnormalformphase} recovers the $r$-column.
This transformation is shown in Figure~\ref{fig:phasetransformationtableau}.
In our tool, we apply the recovery algorithm after synthesizing the optimal circuit.
After every synthesis run, we compare the tableau of input and output circuits for correctness.

\begin{algorithm}[htbp]
\caption{Relative Phase Recovery}
\label{alg:recovery}
\begin{algorithmic}[1]
\STATE SEQ is empty
\FORALL{$i \in [0 \dots n-1]$}
\IF{$r_i \oplus r'_i = 1$}
\STATE APPEND $\text{Z}_i$ to SEQ
\ENDIF
\IF{$r_{n+i} \oplus r'_{n+i} = 1$}
\STATE APPEND $\text{X}_i$ to SEQ
\ENDIF
\ENDFOR
\end{algorithmic}
\end{algorithm}

\section{CNOT Optimal SAT Encodings}
\label{sec:satencodings}
We now present bounded reachability SAT encodings for cx-count and
cx-depth optimization based on the ideas discussed above.
For cx-count optimization, we need to reach the target tableau in $d$ entangling steps from the initial tableau.
Encoding tableau updates for the entangling sequences in one step $t$ to $t+1$ is quite complex.
Instead, we define a \emph{layer} that acts on $t, t+1, t+2$ time steps.
We encode 1-qubit gate updates from $t$ to $t+1$ and CNOT updates from $t+1$ to $t+2$.
In each layer, we choose a control and target qubit on which the entangling sequence is applied.
At the end, we need to apply a last layer of all $6$ 1-q sequences.
In total, we need $2d+2$ time steps i.e., from $t=0$ to $t=2d+1$ for a $d$-CNOT circuit encoding.
Algorithm~\ref{alg:encodingoutline} presents the overall outline of the SAT encoding for a $d$-CNOT, $\numqubits$-qubit circuit.
For cx-depth optimization, instead of exactly one control and target, we will allow any subset of independent (i.e., parallel) qubit pairs.
Essentially, optimizing for the number of layers corresponds to optimizing for the chosen cx-count or cx-depth metric.

\begin{algorithm}[htbp]
\caption{SAT Encoding Outline for $d$-CNOT Circuit}
\label{alg:encodingoutline}
\begin{algorithmic}[1]
\STATE InitialStateConstraints \COMMENT{Encode initial tableau at $t=0$ as in~\ref{subsubsec:intialgoalconstraints}}
\STATE GoalStateConstraints    \COMMENT{Encode target tableau at $t=2d+1$ as in~\ref{subsubsec:intialgoalconstraints}}
\FORALL{$k \in [0 \dots d-1]$}
\STATE Encode1qConstraints($k$)  \COMMENT{Encode I,HS,SH constraints in $k$th layer as in~\ref{subsubsec:singleconstraints}}
\STATE EncodeCNOTConstraints($k$)\COMMENT{Encode CNOT constraints in $k$th layer as in~\ref{subsubsec:cnotconstraints}}
\ENDFOR
\STATE Last1qConstraints       \COMMENT{Encode last 1q constraints at $t=2d$ as in~\ref{subsubsec:finallayer}}
\end{algorithmic}
\end{algorithm}

\subsection{Gate optimal encoding}
\label{subsec:gateoptimal}

\begin{table}[htbp]
\centering
\caption{Encoding variables and descriptions}
\begin{tabular}{ll}
\toprule
Variable & Description \\
\midrule
$\ivar^{k}_{a}/\hsvar^{k}_{a}/\shvar^{k}_{a}$ & apply I/HS/SH gate on $i$th qubit in the layer $k$\\
$\cnotvar^{k}_{a,b}$ & apply CNOT gate on $a,b$ qubits ($a<b$) in the layer $k$\\
$\ctrlvar^{k}_a/\trgtvar^{k}_b$ & choose $a$ as a control/target qubit in the layer $k$\\
$\ivar^{\last}_{a}/\hvar^{\last}_{a}/\svar^{\last}_{a}/\hsvar^{\last}_{a}/\shvar^{\last}_{a}/\hshvar^{\last}_{a}$
& apply I/H/S/HS/SH/HSH gates on $a$th qubit in the last layer\\
\cmidrule(lr){1-2}
$\xmatrixvar_{i,a}^{t}/\zmatrixvar_{i,a}^{t}$ & state of $x_{ia}/z_{ia}$ matrix element at time step $t$\\
$\px_{i,a}^{t}/\pz_{i,a}^{t}$ & propagate $x_{ia}/z_{ia}$ matrix element from time step $t$ to $t+1$\\
\bottomrule
\end{tabular}
\label{table:vars}
\end{table}

We define the variables as shown in Table~\ref{table:vars}.
The variables in top half represent chosen control and target qubits, and applied gates.
Variables in bottom half represent states of matrix elements.
In addition, we define auxiliary variables to encode XOR constraints elegantly.
We essentially represent propagation of $x,z$ matrix states from $t$ to $t+1$ time steps.
For time steps $t=0$ to $t=2k$, we first encode propagation constraints:

\begin{align}
\bigwedge_{a=0}^{\numqubits-1} \bigwedge_{i=0}^{2\numqubits-1}
\big((\px^{t}_{i,a} \iff (\xmatrixvar_{i,a}^{t} \iff \xmatrixvar_{i,a}^{t+1}))
\land (\pz^{t}_{i,a} \iff (\zmatrixvar_{i,a}^{t} \iff \zmatrixvar_{i,a}^{t+1}))\big)
\end{align}
In the following Subsections, we present constraints for each macro in Algorithm~\ref{alg:encodingoutline}.

\subsubsection{Encoding 1-q I, HS, SH constraints}
\label{subsubsec:singleconstraints}

For a given layer number $k$, we define $t := 2k$.
Exactly one of the 3 single qubit gates is chosen on each qubit.
We apply sequential counter ExactlyOne constraints from PySAT~\cite{imms-sat18}, referred as EO from here on.
If a qubit is neither a control nor target, then we apply identity gate.
Instead, if HS or SH gate is applied, the qubit must be either a control or target.
\begin{equation}
\bigwedge_{a=0}^{\numqubits-1} \EO(\ivar^{k}_{a}, \hsvar^{k}_{a}, \shvar^{k}_{a}) \land
\bigwedge_{a=0}^{\numqubits-1} (\neg \ctrlvar^{k}_a \land \neg \trgtvar^{k}_a) \implies \ivar^{k}_a \land
\bigwedge_{a=0}^{\numqubits-1} ( \hsvar^{k}_a \lor \shvar^{k}_a) \implies (\ctrlvar^{k}_a \lor \trgtvar^{k}_a)
\end{equation}
For each of the three gates {I, HS, SH}, we encode following constraints:
\begin{align}
&\bigwedge_{a=0}^{\numqubits-1} \bigwedge_{i=0}^{2\numqubits-1} \big( &\\
&\quad\ivar^{k}_a \implies (\px^{t}_{i,a} \land \pz^{t}_{i,a})
& \triangleright\text{I: } x_{ia} := x_{ia};\text{ } z_{ia} := z_{ia}\quad\quad\quad\text{ }\nonumber\\
&\quad\hsvar^{k}_a \implies
  \big((\zmatrixvar_{i,a}^{t} \iff \xmatrixvar_{i,a}^{t+1}) \land (\xmatrixvar_{i,a}^{t} \iff \neg\pz_{i,a}^{t})\big)
  & \triangleright\text{HS: } x_{ia} := z_{ia};\text{ }  z_{ia} := x_{ia} \oplus z_{ia} \nonumber\\
&\quad\shvar^{k}_a \implies
  \big((\xmatrixvar_{i,a}^{t} \iff \zmatrixvar_{i,a}^{t+1}) \land (\zmatrixvar_{i,a}^{t} \iff \neg\px_{i,a}^{t})\big)\big)
  & \triangleright\text{SH: } z_{ia} := x_{ia};\text{ }  x_{ia} := z_{ia} \oplus x_{ia}\nonumber
\end{align}

\subsubsection{Encoding CNOT constraints}
\label{subsubsec:cnotconstraints}
Let $k$ be the layer number, then we define $t := 2 \cdot k + 1$.
We schedule exactly one CNOT gate in every layer and
chosen CNOT defines the control and target qubits.
\begin{equation}
\EO(\{{\cnotvar^{t}_{a,b}\mid 0 \leq a < \numqubits; a < b < \numqubits}\}) \land
\bigwedge_{a=0}^{\numqubits-1}\bigwedge_{b=a+1}^{\numqubits-1} {\cnotvar^{t}_{a,b} \iff (\ctrlvar^{t}_{a} \land \trgtvar^{t}_{b})}
\end{equation}
Similar to 1q constraints, we define the tableau updates rules as below.
\begin{align}
  \bigwedge_{a=0}^{\numqubits-1}\bigwedge_{b=a+1}^{\numqubits-1}\bigwedge_{i=0}^{2\numqubits-1}
          \cnotvar^{k}_{a,b}& \implies \big( \quad\quad\quad\quad\quad\quad\quad\quad\quad\quad
          \triangleright  x_{ib} := x_{ia} \oplus x_{ib} ;\text{ }  z_{ia} := z_{ib} \oplus z_{ia}\nonumber\\
          &(\xmatrixvar_{i,a}^{t} \iff \neg\px_{i,b}^{t}) \land (\zmatrixvar_{i,b}^{t} \iff \neg\pz_{i,a}^{t})\big) 
\end{align}
\begin{align}
\bigwedge_{a=0}^{\numqubits-1} \bigwedge_{i=0}^{2\numqubits-1} (\neg \trgtvar^{k}_a \implies \px_{i,a}^{t})
\land \bigwedge_{b=0}^{\numqubits-1} \bigwedge_{i=0}^{2\numqubits-1}(\neg \ctrlvar^{k}_b \implies \pz_{i,b}^{t})
& \quad\text{ }\triangleright x_{ia} := x_{ia};\text{ } z_{ib} := z_{ib}
\end{align}
Given a coupling, for every non-neighbor $(a,b)$ qubit pair we add $\neg \cnotvar^{t}_{a,b}$.
\subsubsection{Final Layer 1q Constraints}
\label{subsubsec:finallayer}
We define $t=2d$ and apply exactly one of the $6$ 1-q unique gate sequences for each qubit $a$
as $\bigwedge_{a=0}^{\numqubits-1} \EO(\ivar^{\last}_a, \hvar^{\last}_a, \svar^{\last}_a, \hsvar^{\last}_a, \shvar^{\last}_a, \hshvar^{\last}_a)$.
The constraints for gate sequences \{I, HS, SH\} are exactly same as transition constraints
presented earlier as in~\ref{subsubsec:singleconstraints}.
We now give constraints for rest of the three gate sequences.
\begin{align}
\bigwedge_{i=0}^{\numqubits-1} &\bigwedge_{r=0}^{2\numqubits-1} \big(\\
&\hvar^{\last}_{i} \implies \quad
  \big((\zmatrixvar_{i,a}^{t} \iff \xmatrixvar_{i,a}^{t+1}) \land (\xmatrixvar_{i,a}^{t} \iff \zmatrixvar_{i,a}^{t+1})\big)
  &\triangleright\text{H: } x_{ia} := z_{ia};\text{ } z_{ia} := x_{ia}\text{ }\quad\quad\nonumber\\
&\svar^{\last}_{i} \implies \quad
  \big(\px_{i,a}^{t} \land \text{ } (\xmatrixvar_{i,a}^{t} \iff \neg\pz_{i,a}^{t})\big)
  &\triangleright\text{S: } x_{ia} := x_{ia};\text{ }  z_{ia} := x_{ia} \oplus z_{ia}\nonumber\\
&\hshvar^{\last}_{i} \implies
\big(\pz_{i,a}^{t} \land \text{ } (\zmatrixvar_{i,a}^{t} \iff \neg\px_{i,a}^{t})\big)\big)
& \triangleright\text{HSH: } z_{ia} := z_{ia};\text{ }  x_{ia} := z_{ia} \oplus x_{ia} \nonumber
\end{align}

\subsubsection{Initial and Goal Constraints}
\label{subsubsec:intialgoalconstraints}

Recall our discussion on tableau structure and initial tableau in Section~\ref{subsec:cliffordcircuits}.
Let us divide the tableau without r-column into 4 equal quadrants.
The left quadrants corresponds to the x matrix and the right quadrants correspond to the z matrix,
where columns are labelled with qubits.
Initially, the top-left and bottom-right quadrants correspond to the Identity matrix, whereas the other quadrants are zero matrices.
Permuting columns of x and z matrices corresponds to qubit relabeling.
Exactly-One constraints on rows and column of top-left quadrant represents column permutation.
\begin{align}
\bigwedge_{i=0}^{n-1}\big(&\bigwedge_{a=0}^{n-1} \big((\xmatrixvar_{i,a}^{0} \iff \zmatrixvar_{i,n+a}^{0})
\land \neg \xmatrixvar_{n+i,a}^{0} \land \neg \zmatrixvar_{i,a}^{0}\big) \land \nonumber \\
&\EO(\{\xmatrixvar_{i,a}^{0}\mid 0 \leq a < n\}) \land \EO(\{\xmatrixvar_{i,a}^{0}\mid 0 \leq a < n\})\big)
\end{align}
We can disable column permutation (qubit relabeling) by forcing top-left quadrant as the Identity matrix
i.e., adding clauses $\bigwedge_{i=0}^{n-1} \xmatrixvar_{i,i}^{0}$.
For goal constraints, based on the target tableau we set the final state variables.
For each $x_{ia}$ element, we add unit clause $\xmatrixvar_{i,a}^{2d+1}$
if $x_{ia}=1$ else we add $\neg \xmatrixvar_{i,a}^{2d+1}$.
We add similar clauses for all $z$ matrix elements.

\subsubsection{Improvements due to Search Space Reduction}
\label{subsubsec:improvements}

We observe that the order between two parallel entangling sequences does not matter.
Let $\cnotvar_{a,b}^{k-1}$ and $\cnotvar_{a',b'}^{k}$ be two parallel gates, i.e., they do not share control or target qubits.
We order them based on the control qubits, i.e., if $a > a'$, then we add $\neg \cnotvar_{a,b}^{k-1} \lor \neg\cnotvar_{a',b'}^{k}$.
Note that gate ordering constraints increases the number of clauses from $O(d n^3)$ to $O(d n^4)$.
Yet, these extra binary clauses reduce the search space and help the SAT solver.

In an optimal makespan, a valid solution can never repeat the state in any layer i.e., we need to consider simple paths only.
We can break any $m$-cycle by encoding that every layer pair $k, k'$ is different in at least one state variable.
Essentially, given layers $k,k'$, we use indicator variables $\dr_{a}^{k,k'}, \dxc_{a}^{k,k'}, \dzc_{a}^{k,k'}$
representing which row and x-column or z-column is different, respectively.
The following constraints encode that if the $i$th row and the $a$th column is chosen, then the corresponding matrix variable is different.
We need to find exactly one such matrix variable which is different.
\begin{equation*}
\bigwedge_{a=0}^{\numqubits-1} \bigwedge_{i=0}^{2\numqubits-1} ((\dr_{i}^{k,k'} \land \dxc_{a}^{k,k'})
\implies (\xmatrixvar_{i,a}^{2k+2} \neq \xmatrixvar_{i,a}^{2k'+2})) \land ((\dr_{i}^{k,k'} \land \dzc_{a}^{k,k'})
\implies (\zmatrixvar_{i,a}^{2k+2} \neq \zmatrixvar_{i,a}^{2k'+2}))\\
\end{equation*}
\begin{equation*}
\land \EO(\{\dr_{i}^{k,k'}\mid 0 \leq i < 2\numqubits\}) \land \EO(\{\dxc_{i}^{k,k'} \mid 0 \leq i < \numqubits\}
\cup \{\dzc_{i}^{k,k'} \mid 0 \leq i < \numqubits\})
\end{equation*}
By default, we break all $3$-cycles, since it doesn't increase the asymptotic variable and clause count.
Further, we optionally allow some redundant clauses and auxiliary variables as
in \cref{sec:redundantclauses} to help our backend SAT solver.

\subsection{Depth optimal encoding}
\label{subsec:depthoptimal}

With minimal changes, we can encode optimal depth instead of optimal gate count.
Instead of allowing a single CNOT gate, we allow multiple independent gates in each layer.
In other words, we allow at most one CNOT gate at any control or target qubit.
We essentially replace Equation 4 with following Equation 9.
\begin{align}
& \bigwedge_{q=0}^{\numqubits-1} \AMO(\{\cnotvar_{a,b}^{k} \mid a=q \text{ or } b=q \text{ and } 0 \leq a,b < n \})
\land \ALO(\{\cnotvar_{a,b}^{t} \mid 0 \leq a,b < n\}) \land\nonumber\\
& \bigwedge_{q=0}^{\numqubits-1} \big((\ctrlvar^{k}_{q} \implies \bigvee_{b=0}^{\numqubits-1}\cnotvar_{q,b}^{k})
\land (\trgtvar^{k}_{q} \implies \bigvee_{a=0}^{\numqubits-1}\cnotvar_{a,q}^{k})\big)
\end{align}
Gate ordering restrictions are simpler for Depth optimal encoding.
We eagerly schedule entangling gates without losing optimality.
If a CNOT is applied on qubits $i,j$, we restrict that some entangling gate is applied on $i$ or $j$ qubits in the previous layer.
Note that we only require $O(d n^3)$ clauses even with the additional gate ordering constraints.
\begin{equation*}
\bigwedge_{i=0}^{\numqubits-1} \bigwedge_{j=i+1}^{\numqubits-1} \cnotvar_{i,j}^{k} \implies
\big(\ctrlvar^{k-1}_{i} \lor \ctrlvar^{k-1}_{j} \lor  \trgtvar^{k-1}_{i} \lor \trgtvar^{k-1}_{j}\big)
\end{equation*}

\section{Experiments and Results}
\label{sec:experiments}

We implemented above encodings and their variations in version 5 of our open source tool Q-Synth.
Given a quantum circuit in OPENQASM 2.0~\cite{cross2017openquantumassemblylanguage} format,
we return the optimized circuit in the same format.
For \emph{pure} Clifford circuits, we return CNOT-optimal circuits for a given metric.
Given a mapped circuit and a coupling graph of a quantum platform, we optimize respecting the coupling restrictions.
Similarly, we generate circuits with a possible qubit relabeling when enabled.
On the other hand, we employ peephole synthesis for arbitrary circuits.
From start to end of the circuit, we first group non-Clifford and Clifford gates greedily.
Then we replace each obtained Clifford slice by its optimized counterpart.
Note that for general circuits this does not guarantee global optimality either in cx-count or cx-depth optimization.

\paragraph*{Search Strategies}
For each Clifford optimization call, we provide two search strategies i.e., forward and backward.
In forward search, we search for circuits with $k$ CNOT gates, increasing from $k=0$ until we find the optimal circuit.
Forward search either produces an optimal circuit or a timeout.
In backward search, we search for circuits with $\leq k$ CNOT gates, decreasing from the number of CNOT gates in the initial circuit until we reach UNSAT.
Within the given time limit, we either synthesize an optimal circuit or report the best circuit found so far.
In our initial experiments, we observed that search space reduction techniques help forward search.
Below, per experiment we only present the best combination we observed.
For cx-count optimization, we also include gate ordering constraints.
For cx-depth optimization, we include both gate ordering constraints and simple path restrictions.

\paragraph*{Research Questions}
We are interested in investigating the effectiveness and relevance of our SAT based approach.
In particular, we are interested in answering the following three research questions:
\begin{itemize}
\item R1: Does restricting search to Clifford normal forms help with scalability?
\item R2: What is the effectiveness of SAT encodings on unmapped practical benchmarks?
\item R3: What is the effectiveness of SAT encodings on layout mapped practical benchmarks?
\end{itemize}
We propose the following 3 experiments to answer the above research questions.

\subsection{Experiment 1: Optimal synthesis}
The main difference from existing SAT based approaches is our use of normal forms.
In this experiment, we investigate if normal form restricted search helps with scalability.
For comparison, we consider QMAP~\cite{Wille_2023} tool that implements previous exact sat-based synthesis of Clifford circuits.
QMAP provides near-optimal synthesis for cx-count~\cite{Schneider2022ASE} and optimal synthesis for circuit depth~\cite{Peham2023DepthOptimalSO}.
In~\cite{Schneider2022ASE,Peham2023DepthOptimalSO}, QMAP presented results mainly on synthesis of random Clifford circuits from given stabilizers.
Random Clifford circuits are one of the standard benchmarks for comparing various approaches.
Unfortunately, the random stabilizers used by QMAP in ~\cite{Schneider2022ASE,Peham2023DepthOptimalSO} are not accessible.
Thus, we generated new random benchmarks of 3 to 7 qubits, with 5 random instances for each qubit.
For benchmark generation, we used the standard Qiskit random Clifford stabilizer function.
For each such stabilizer, we further optimized with TKET~\cite{Sivarajah_2021} both with and without permutation.
Our input benchmarks are essentially the best circuits synthesized by Qiskit and TKET.
Any improvement on such benchmarks essentially shows the strength of exact approaches.
All generated benchmarks and their scripts are available online (for reproducibility).

For cx-count optimization, we compared with near-optimal cx-count optimization by QMAP~\cite{Schneider2022ASE}.
For cx-depth optimization, we compare with optimal circuit depth optimization by QMAP~\cite{Peham2023DepthOptimalSO}
and report cx-depth instead of circuit depth.
We present results for Q-Synth with forward and backward search, with- and without-permutations.
For each instance, we give 3hr time limit and 8GB memory limit. For Q-Synth, we use state-of-the-art SAT solver Cadical (v2.1)~\cite{BiereFazekasFleuryHeisinger-SAT-Competition-2020-solvers} as a backend.
We use Pysat~\cite{imms-sat18} for sequential counter cardinality constraints.
\begin{table}[htbp]
  \caption{Experiment 1: Optimizing of random Clifford circuits, n: qubits, m: optimization metric (average cx-count/cx-depth of 5 instances),
  t: average time if all 5 instances are completely solved; else the number of instances solved within the time limit.}
  \label{tb:experiment1}
  \centering
  \begin{tabular}{clrrrrrrrrrrrr}
  \toprule
  && \multicolumn{7}{c}{without-permutation} & \multicolumn{5}{c}{with-permutation}\\
  \cmidrule(lr){3-9} \cmidrule(lr){10-14}
  &      & tket & \multicolumn{2}{c}{qmap} & \multicolumn{2}{c}{qsynth} &\multicolumn{2}{c}{qsynth-b} & tket & \multicolumn{2}{c}{qsynth} &\multicolumn{2}{c}{qsynth-b}\\
  \cmidrule(lr){4-5} \cmidrule(lr){6-7} \cmidrule(lr){8-9} \cmidrule(lr){11-12} \cmidrule(lr){13-14}
  &n & m & m &t & m & t & m & t & m & m & t & m & t \\
  \cmidrule(lr){1-14}
  \multirow{5}{*}{\rotatebox[origin=c]{90}{\shortstack{cx-count}}}&3 & 3.6 & 3.6 & 1002.8 & 3.6 & 0.1 & 3.6 & 0.1 & 2.8 & 2.8 & 0.1 & 2.8 & 0.1 \\
  &4 & 6.8 & - & 0/5 & 6.0 & 0.3 & 6.0 & 0.2 & 6.0 & 4.8 & 0.3 & 4.8 & 0.4 \\
  &5 & 13.2 & - & 0/5 & 8.2 & 7.1 & 8.2 & 21.9 & 10.2 & 7.2 & 19.5 & 7.2 & 70.3 \\
  &6 & 17.0 & - & 0/5 & 12.6 & 3/5 & 11.6 & 2/5 & 14.4 & 13.6 & 1/5 & 11.0 & 0/5 \\
  &7 & 22.8 & - & 0/5 & - & 0/5 & 21.0 & 0/5 & 20.4 & - & 0/5 & 18.4 & 0/5 \\
  \cmidrule(lr){1-14}
  \multirow{5}{*}{\rotatebox[origin=c]{90}{\shortstack{cx-depth}}}&3 & 3.6 & 3.6 & 0.4 & 3.6 & 0.1 & 3.6 & 0.1 & 2.8 & 2.8 & 0.1 & 2.8 & 0.1 \\
  &4 & 5.6 & 4.8 & 5.1 & 3.6 & 0.1 & 3.6 & 0.2 & 4.8 & 3.0 & 0.1 & 3.0 & 0.2 \\
  &5 & 11.2 & 7.2 & 434.3 & 4.8 & 1.2 & 4.8 & 16.2 & 8.2 & 4.0 & 2.5 & 4.0 & 5.8 \\
  &6 & 13.6 & 13.2 & 0/5 & 5.0 & 10.5 & 5.0 & 839.4 & 11.2 & 5.0 & 82.3 & 5.0 & 583.0 \\
  &7 & 18.0 & - & 0/5 & 8.6 & 4/5 & 17.8 & 0/5 & 15.4 & - & 0/5 & 15.0 & 0/5 \\
  \bottomrule
  \end{tabular}
  \end{table}
\paragraph*{Results and Discussion}
Table~\ref{tb:experiment1} presents the results of Experiment 1 with cx-count and cx-depth optimization.
For cx-count optimization, we can see that the near-optimal cx-count synthesis by QMAP does not scale well
and only solves 3-qubit instances.
Q-Synth, on the other hand, with forward search can solve all instances up to 5-qubits and 3 out of 5 6-qubit instances.
With backward search, Q-Synth even improves cx-count of 7-qubit instances but uses full 3 hour time limit to do so.
In the presence of output permutation, the results are similar to without output permutation while solving fewer 6-qubit instances.
For cx-depth optimization, QMAP (without output permutation) can now solve up to 5-qubit instances with optimal depth.
Note that QMAP often reports suboptimal cx-depth when optimizing depth.
For the hard 6-qubit instances QMAP produces intermediate results, but they are far from optimal.
Q-Synth on the other hand, solves all but one 7-qubit instance with forward search.
Interestingly with backward search we can improve the unsolved 7-qubit instance but not solve optimally.
With output permutation, Q-Synth can optimally solve up to 6-qubit instances but not 7-qubit instances.
With backward search, Q-Synth can improve 1 extra 7-qubit instance.
We provide the full data in Tables~\ref{tb:experiment1cxc} and~\ref{tb:experiment1cxd} in Appendix.

From the results, it is clear that restricting search to normal form indeed helps with scalability.
We outperform previous SAT approaches by several orders of magnitude in both optimization metrics.
Notice that restricting to normal forms achieves two goals.
First, the makespan in our encodings correspond to cx-count or cx-depth.
On the other hand, the makespan in previous encodings correspond to gate count and circuit depth.
Note that gate count or circuit depth can be several times higher than their counterparts.
Second, we avoid search through many equivalent circuits thus reducing search space.
We conjecture that good performance is due to lower makespan and reduced search space.

Coming to weaknesses of our approach, we could not solve all $6$-qubit instances for cx-count optimization.
Existing exhaustive based approach~\cite{Bravyi_2022} could generate a database of all $6$-qubit instances.
Essentially, an exhaustive approach allows more freedom in avoiding search of equivalent (or symmetric) circuits.
It is unclear to us how to include such symmetry breaking in a SAT encoding.
On the other hand, our cx-depth optimization can go beyond 6-qubits which is not feasible in exhaustive approach.

\subsection{Experiment 2: Effect on All-to-All Practical Benchmarks}
In practical benchmarks, the Clifford sub-circuits are often shallow.
We are interested in investigating the effect of our SAT based optimization on such benchmarks.
Here we optimize benchmarks assuming all-to-all qubit connectivity.
For practical benchmarks, we consider standard variational quantum eigensolver (VQE) benchmarks as the first benchmark set.
Similar to~\cite{shaik2024optimal}, we generated 12 random VQE benchmarks of 8 and 16 qubits with up to 536 cx-depth and 860 cx-count.
For the second benchmark set, we used 28 instances from the standard benchmark collection of Feynman tool~\cite{feynman2016} with up to
24 qubits, 2149 cx-count, and 1878 cx-depth.
We consider best cx-count reduction compiler TKET~\cite{Sivarajah_2021, nation2025benchmarkingperformancequantumcomputing} for a comparison.
We also experiment with combined optimization of TKET and Q-Synth, we denote it by TK+QS.
In the TK+QS configuration, we first optimize each instance with TKET and then optimize with Q-Synth.
In the experiment with qubit permutation, we also enable permutation in TKET.
Waiting 3 hours for a single benchmark is not practical.
In Experiments 2 and 3, we run Q-Synth only with forward search, with a 600s time limit.
\begin{table}[htbp]
  \caption{Experiment 2: Optimizing practical benchmarks; org: original optimization metric, ch\%: change\% of optimization metric (lower is better), t: time.}
  \label{tb:experiment2}
  \centering
  \resizebox{\columnwidth}{!}{%
  \begin{tabular}{clrrrrrrrrrrr}
  \toprule
  && & \multicolumn{5}{c}{without-permutation} & \multicolumn{5}{c}{with-permutation}\\
  \cmidrule(lr){4-8} \cmidrule(lr){8-13}
  && & tket & \multicolumn{2}{c}{qsynth} & \multicolumn{2}{c}{TK+QS} & tket & \multicolumn{2}{c}{qsynth} & \multicolumn{2}{c}{TK+QS}\\
  \cmidrule(lr){5-6} \cmidrule(lr){7-8} \cmidrule(lr){10-11} \cmidrule(lr){12-13}
  && org & ch\% & ch\% & t & ch\% & t & ch\% & ch\% & t & ch\% & t \\
  \cmidrule(lr){1-13}
  \cmidrule(lr){3-7} \cmidrule(lr){8-12}
  \multirow{3}{*}{\rotatebox[origin=c]{90}{\shortstack{VQE \\ cx-count}}}&avg & 575.8 & -3.5 & -3.3 & 321.9 & -3.9 & 318.4 & -5.8 & -4.4 & 371.4 & -6.1 & 364.0 \\
  &min & 298 & -7.7 & -8.2 & 8.3 & -8.5 & 8.2 & -8.5 & -8.9 & 16.4 & -9.8 & 16.2 \\
  &max & 860 & -0.6 & -0.6 & 604& -0.6 & 604.4 & -0.7 & -0.7 & 603.2 & -0.7 & 603.3 \\
  \cmidrule(lr){1-13}
  \cmidrule(lr){3-7} \cmidrule(lr){8-12}
  \multirow{3}{*}{\rotatebox[origin=c]{90}{\shortstack{VQE \\ cx-depth}}}&avg & 367.6 & -2.9 & -9.1 & 20.1 & -9.6 & 19.2 & -4& -10.8 & 72.2 & -11.2 & 58.8 \\
  &min & 233 & -8.2 & -16& 3.1 & -16.7 & 3.4 & -8.6 & -16.7 & 4.9 & -17.1 & 4.8 \\
  &max & 536 & 0& -3& 60.5 & -3.8 & 51.9 & 0& -3.8 & 363.5 & -4.7 & 230.6 \\
  \cmidrule(lr){1-13}
  \multirow{3}{*}{\rotatebox[origin=c]{90}{\shortstack{Feynman \\ cx-count}}}&avg & 222& -8.7 & -8.4 & 207.2 & -11.2 & 208.7 & -9.8 & -8.8 & 233.3 & -12.4 & 218.4 \\
  &min & 18 & -28.6 & -32.1 & 0.2 & -32.1 & 0.2 & -28.6 & -32.1 & 0.2 & -32.6 & 0.2 \\
  &max & 2149 & 0& 0& 612.3 & 0& 613.5 & 0& 0& 610.3 & 0& 609.5 \\
  \cmidrule(lr){1-13}
  \multirow{3}{*}{\rotatebox[origin=c]{90}{\shortstack{Feynman \\ cx-depth}}}&avg & 161& -7.8 & -12.3 & 89.7 & -14.4 & 89.5 & -8.3 & -12.6 & 104.5 & -14.7 & 104.2 \\
  &min & 16& -29.6 & -48.1 & 0.2 & -48.1 & 0.2 & -29.6 & -48.1 & 0.2 & -48.1 & 0.2 \\
  &max & 1878 & 0& 1.9 & 604.3 & 0& 604.2 & 0& 4.9 & 603.8 & 1.9 & 603.7 \\
  \bottomrule
  \end{tabular}
  }
  \end{table}
\paragraph*{Results and Discussion}
Tables~\ref{tb:experiment2} present results of Experiment 2.
We present average, minimum, and maximum change in optimization metric and time taken.
For cx-count optimization, Q-Synth produces similar reductions compared to TKET in both benchmark sets.
We observed up to 8.9\% and 32.1\% reduction in VQE and Feynman benchmarks respectively by Q-Synth.
There does exist some hard slices where Q-Synth timeouts resulting in better results by TKET.
For cx-depth reduction, Q-Synth overall outperforms TKET for both benchmarks sets.
We observed up to 16.7\% and 48.1\% reduction in VQE and Feynman benchmarks respectively by Q-Synth.
We usually observe the best results with TKET+Q-Synth combination.
Tables~\ref{tb:experiment2a_vqe_cxc} to~\ref{tb:experiment2b_feynman_cxd},
in the Appendix provide the full data for all variations.

From the results, we can see that SAT based approaches perform well also on practical benchmarks.
Remember that our benchmarks can have up to 24 qubits, 2149 cx-count, and 1878 cx-depth.
Q-Synth produces better cx-depth results up to 48.1\% compared to 29.1\% by TKET.
Such reductions are feasible since TKET focuses on cx-count.
Indeed, TKET on an average produces better cx-count reduction than Q-Synth on large instances.
Since cx-count and cx-depth are both important metrics, optimizing both is helpful in practice.
We see the best results by combining the strengths of TKET and Q-Synth in TK+QS combination.
Our SAT based cx-depth optimization appears to nicely complement heuristic cx-count optimization.
We conclude that using the combination TK+QS for reduction in both cx-count and cx-depth is feasible in a practical compilation setting.

\subsection{Experiment 3: Effect on Mapped Practical Benchmarks}

Layout synthesis routines in industrial compilers are fast but sub-optimal.
Many Clifford optimization techniques do not support layout-aware synthesis, for example TKET.
Intuitively, connectivity restrictions break assumptions made in all-to-all connectivity.
SAT based approaches can elegantly encode connectivity constraints.
In this experiment, we investigate if our approach can further reduce mapped practical circuits by industrial compilers.
Our goal is to investigate the reduction in best synthesized benchmarks from industrial compilers.
We consider TKET optimized benchmarks with qubit permutations from Experiment 2.
We then mapped the benchmarks onto 54-qubit Sycamore~\cite{arute2019quantum},
80-qubit Rigetti~\cite{computingrigetti} and 127-qubit Eagle~\cite{chow2021ibm} platforms using Qiskit.
During Qiskit transpiling, we used the highest optimization "-O3" level.
For each benchmark, we optimize with Q-Synth for both cx-count and cx-depth.
\begin{table}[htbp]
  \caption{Experiment 3: Optimizing mapped practical benchmarks; org: original optimization metric,
  ch\%: change\% of optimization metric (lower the better), t: time.}
\label{tb:experiment3}
\centering
\begin{tabular}{clrrrrrrrrr}
\toprule
&& \multicolumn{3}{c}{sycamore-54}& \multicolumn{3}{c}{rigetti-80}& \multicolumn{3}{c}{eagle-127}\\
\cmidrule(lr){3-5} \cmidrule(lr){6-8} \cmidrule(lr){9-11}
& & org & ch\%  & t & org & ch\%  & t & org  & ch\%  & t \\
\cmidrule(lr){1-11}
\multirow{3}{*}{\rotatebox[origin=c]{90}{\shortstack{VQE \\ cx-count}}}&avg & 1253.7 & -7.7 & 604.1 & 1550.2 & -8.8 & 603.8 & 1842.9 & -9.4 & 568\\
&min & 515& -15.9 & 602.2 & 644& -19.3 & 602.7 & 731& -18.3 & 158.1 \\
&max & 2179& -2& 607.4 & 2698& -2.3 & 605.1 & 3172& -2.5 & 607.2 \\
\cmidrule(lr){1-11}
\multirow{3}{*}{\rotatebox[origin=c]{90}{\shortstack{VQE \\ cx-depth}}}&avg & 818.1 & -15.8 & 277.9 & 981.3 & -17.2 & 354.2 & 1150.4 & -18.9 & 344.5 \\
&min & 419& -27.4 & 24& 514& -25.9 & 28.5 & 597& -24.8 & 33\\
&max & 1377& -9.1 & 607.8 & 1579& -9.3 & 607.7 & 1886& -12.3 & 611.9 \\
\cmidrule(lr){1-11}
\multirow{3}{*}{\rotatebox[origin=c]{90}{\shortstack{Feynman \\ cx-count}}}&avg & 433.4 & -16.2 & 264.5 & 497.4 & -18.5 & 224.3 & 559.3 & -21& 219\\
&min & 33& -24.8 & 0.7 & 38& -27.3 & 0.5 & 39& -30.3 & 0.6 \\
&max & 3564& -7.9 & 632& 4173& -5.9 & 611.9 & 4809& -9& 622.6 \\
\cmidrule(lr){1-11}
\multirow{3}{*}{\rotatebox[origin=c]{90}{\shortstack{Feynman \\ cx-depth}}}&avg & 308.5 & -22.1 & 97.1 & 349.7 & -22.7 & 119.4 & 386.2 & -23& 134.7 \\
&min & 30& -32.6 & 0.7 & 37& -32.6 & 0.5 & 37& -35.9 & 0.6 \\
&max & 2849& -10.7 & 610.8 & 3322& -7.9 & 610.7 & 3790& -6.6 & 620.9 \\
\bottomrule
\end{tabular}
\end{table}
\paragraph*{Results and Discussion}
Table~\ref{tb:experiment3} present the results for all three platforms.
On all platforms, Q-Synth reports good reduction for both cx-count and cx-depth metrics.
For VQE benchmarks, we observed reduction of up to 19.3\% cx-count and 27.4\% cx-depth.
For Feynman benchmarks, we observed reduction of up to 30.3\% cx-count and 35.9\% cx-depth.
Tables~\ref{tb:experiment3a_vqe_cxc} to~\ref{tb:experiment3b_feynman_cxd},
in the Appendix provide the full data for all variations.

The separation of layout and circuit synthesis in current compilers results in suboptimal circuits.
We see that SAT based approaches are effective in post layout optimization.
From the full tables, we also observe that mapping and optimizing to Google's Sycamore produces the best results.
We conjecture that the denser array structure of the coupling graph is the main factor.
The coupling graphs of Rigetti with octagons and Eagle with heavy-hex require many extra CNOT gates to route qubits, at least on our benchmarks.
Despite our good reductions, the final cx-count/cx-depth on these
platforms are higher than on Sycamore.
Overall, the results indicate that SAT based optimization is useful in the compilation pipeline.

\section{Conclusion}
\label{sec:conclusion}

In this paper, we proposed two SAT encodings that synthesize Clifford circuits with optimal CNOT count or depth.
We also handle connectivity restrictions and allow qubit permutations.
We implemented all encoding variations in the open source tool Q-Synth.
The experiments demonstrate scalability compared to existing SAT encodings.
For the first time, we are able to synthesize cx-depth optimal 7-qubit random Clifford circuits.
On practical VQE and Feynman benchmarks, we overall perform better than TKET in all-to-all connectivity.
We also investigated layout-aware optimization on major quantum platforms.
We first optimize benchmarks with TKET, then map them using Qiskit and 
finally optimize the results using Q-Synth.
This achieves reductions of up to 30.3\% in cx-count and 35.9\% in cx-depth.
These results show that our SAT based approach nicely complements existing heuristic approaches.

\bibliography{clifford_synthesis}

\begin{thebibliography}{10}

\bibitem{Aaronson_2004}
Scott Aaronson and Daniel Gottesman.
\newblock Improved simulation of stabilizer circuits.
\newblock {\em Physical Review A}, 70(5), November 2004.
\newblock \href {https://doi.org/10.1103/physreva.70.052328}
  {\path{doi:10.1103/physreva.70.052328}}.

\bibitem{feynman2016}
Matthew Amy.
\newblock Quantum circuit analysis toolkit, 2016.
\newblock URL: \url{https://github.com/meamy/feynman}.

\bibitem{arute2019quantum}
Frank Arute et~al.
\newblock Quantum supremacy using a programmable superconducting processor.
\newblock {\em Nature}, 574(7779):505--510, 2019.
\newblock \href {https://doi.org/10.1038/s41586-019-1666-5}
  {\path{doi:10.1038/s41586-019-1666-5}}.

\bibitem{BiereFazekasFleuryHeisinger-SAT-Competition-2020-solvers}
Armin Biere, Katalin Fazekas, Mathias Fleury, and Maximilian Heisinger.
\newblock {CaDiCaL}, {Kissat}, {Paracooba}, {Plingeling} and {Treengeling}
  entering the {SAT Competition 2020}.
\newblock In {\em Proc.~of {SAT Competition} 2020 -- Solver and Benchmark
  Descriptions}, volume B-2020-1, pages 51--53. University of Helsinki, 2020.
\newblock URL: \url{https://api.semanticscholar.org/CorpusID:220727106}.

\bibitem{Bravyi_2022}
Sergey Bravyi, Joseph~A. Latone, and Dmitri Maslov.
\newblock 6-qubit optimal clifford circuits.
\newblock {\em npj Quantum Information}, 8(1), July 2022.
\newblock URL: \url{http://dx.doi.org/10.1038/s41534-022-00583-7}, \href
  {https://doi.org/10.1038/s41534-022-00583-7}
  {\path{doi:10.1038/s41534-022-00583-7}}.

\bibitem{chow2021ibm}
Jerry Chow, Oliver Dial, and Jay Gambetta.
\newblock Ibm quantum breaks the 100-qubit processor barrier.
\newblock {\em IBM Research Blog}, 2, 2021.
\newblock URL:
  \url{https://www.ibm.com/quantum/blog/127-qubit-quantum-processor-eagle}.

\bibitem{computingrigetti}
Rigetti Computing.
\newblock Rigetti computing.
\newblock URL: \url{https://www.rigetti.com}.

\bibitem{cross2017openquantumassemblylanguage}
Andrew~W. Cross, Lev~S. Bishop, John~A. Smolin, and Jay~M. Gambetta.
\newblock Open quantum assembly language, 2017.
\newblock URL: \url{https://arxiv.org/abs/1707.03429}, \href
  {http://arxiv.org/abs/1707.03429} {\path{arXiv:1707.03429}}.

\bibitem{DBLP:journals/corr/abs-2205-00724}
Arianne~Meijer{-}van de~Griend and Sarah~Meng Li.
\newblock Dynamic qubit routing with {CNOT} circuit synthesis for quantum
  compilation.
\newblock In {\em Proceedings 19th International Conference on Quantum Physics
  and Logic, {QPL} 2022, Wolfson College, Oxford, UK, 27 June - 1 July 2022},
  volume 394 of {\em {EPTCS}}, pages 363--399, 2022.
\newblock \href {https://doi.org/10.4204/EPTCS.394.18}
  {\path{doi:10.4204/EPTCS.394.18}}.

\bibitem{ettenhuber2024calculatingenergyprofileenzymatic}
Patrick Ettenhuber, Mads~Bøttger Hansen, Irfansha Shaik, Stig~Elkjær
  Rasmussen, Pier~Paolo Poier, Niels~Kristian Madsen, Marco Majland, Frank
  Jensen, Lars Olsen, and Nikolaj~Thomas Zinner.
\newblock Calculating the energy profile of an enzymatic reaction on a quantum
  computer.
\newblock {\em Accepted at Journal of Chemical Theory and Computation}, 2024.
\newblock URL: \url{https://arxiv.org/abs/2408.11091}.

\bibitem{imms-sat18}
Alexey gnatiev, Antonio Morgado, and Joao Marques-Silva.
\newblock {PySAT:} {A} {Python} toolkit for prototyping with {SAT} oracles.
\newblock In {\em SAT}, pages 428--437, 2018.
\newblock \href {https://doi.org/10.1007/978-3-319-94144-8_26}
  {\path{doi:10.1007/978-3-319-94144-8_26}}.

\bibitem{iwama2002transformation}
Kazuo Iwama, Yahiko Kambayashi, and Shigeru Yamashita.
\newblock Transformation rules for designing {CNOT}-based quantum circuits.
\newblock In {\em Proceedings of the 39th annual Design Automation Conference},
  pages 419--424, 2002.

\bibitem{DBLP:conf/soda/JiangSTW0Z20}
Jiaqing Jiang, Xiaoming Sun, Shang{-}Hua Teng, Bujiao Wu, Kewen Wu, and Jialin
  Zhang.
\newblock Optimal space-depth trade-off of {CNOT} circuits in quantum logic
  synthesis.
\newblock In {\em Proceedings of the 2020 {ACM-SIAM} Symposium on Discrete
  Algorithms, {SODA} 2020, Salt Lake City, UT, USA, January 5-8, 2020}, pages
  213--229. {SIAM}, 2020.
\newblock \href {https://doi.org/10.1137/1.9781611975994.13}
  {\path{doi:10.1137/1.9781611975994.13}}.

\bibitem{DBLP:conf/kr/KautzMS96}
Henry~A. Kautz, David~A. McAllester, and Bart Selman.
\newblock Encoding plans in propositional logic.
\newblock In {\em Proceedings of KR-96}, pages 374--384, November 1996.
\newblock URL: \url{https://henrykautz.com/papers/plankr96.pdf}.

\bibitem{OLSQ2_2023}
Wan-Hsuan Lin, Jason Kimko, Bochen Tan, Nikolaj Bjørner, and Jason Cong.
\newblock Scalable optimal layout synthesis for {NISQ} quantum processors.
\newblock In {\em DAC}, 2023.
\newblock URL: \url{https://doi.org/10.1109/DAC56929.2023.10247760}.

\bibitem{Nagarajan2021QuantumCircuitOptAO}
Harsha Nagarajan, Owen Lockwood, and Carleton Coffrin.
\newblock {QuantumCircuitOpt}: An open-source framework for provably optimal
  quantum circuit design.
\newblock {\em 2021 IEEE/ACM Second International Workshop on Quantum Computing
  Software (QCS)}, pages 55--63, 2021.
\newblock URL: \url{https://api.semanticscholar.org/CorpusID:244488668}.

\bibitem{nation2025benchmarkingperformancequantumcomputing}
Paul~D. Nation, Abdullah~Ash Saki, Sebastian Brandhofer, Luciano Bello, Shelly
  Garion, Matthew Treinish, and Ali Javadi-Abhari.
\newblock Benchmarking the performance of quantum computing software.
\newblock {\em CoRR}, abs/2409.08844, 2025.
\newblock \href {http://arxiv.org/abs/2409.08844} {\path{arXiv:2409.08844}}.

\bibitem{Nielsen_Chuang_2010}
Michael~A. Nielsen and Isaac~L. Chuang.
\newblock {\em Quantum circuits}, page 171–215.
\newblock Cambridge University Press, 2010.
\newblock \href {https://doi.org/10.1017/CBO9780511976667.008}
  {\path{doi:10.1017/CBO9780511976667.008}}.

\bibitem{patel2003efficientsynthesislinearreversible}
K.~N. Patel, I.~L. Markov, and J.~P. Hayes.
\newblock Efficient synthesis of linear reversible circuits, 2003.
\newblock URL: \url{https://arxiv.org/abs/quant-ph/0302002}, \href
  {http://arxiv.org/abs/quant-ph/0302002} {\path{arXiv:quant-ph/0302002}}.

\bibitem{Peham2023DepthOptimalSO}
Tom Peham, Nina Brandl, Richard Kueng, Robert Wille, and Lukas Burgholzer.
\newblock Depth-optimal synthesis of {Clifford} circuits with {SAT} solvers.
\newblock {\em 2023 IEEE International Conference on Quantum Computing and
  Engineering (QCE)}, 01:802--813, 2023.
\newblock URL: \url{https://api.semanticscholar.org/CorpusID:258461565}.

\bibitem{Qiskit}
{Qiskit contributors}.
\newblock Qiskit: An open-source framework for quantum computing, 2023.
\newblock \href {https://doi.org/10.5281/zenodo.2573505}
  {\path{doi:10.5281/zenodo.2573505}}.

\bibitem{Schneider2022ASE}
Sarah Schneider, Lukas Burgholzer, and Robert Wille.
\newblock A {SAT} encoding for optimal {Clifford} circuit synthesis.
\newblock {\em 2023 28th Asia and South Pacific Design Automation Conference
  (ASP-DAC)}, pages 190--195, 2022.
\newblock URL: \url{https://api.semanticscholar.org/CorpusID:251800203}.

\bibitem{ShaikvdP2023}
Irfansha Shaik and Jaco van~de Pol.
\newblock Optimal layout synthesis for quantum circuits as classical planning.
\newblock In {\em {IEEE/ACM} International Conference on Computer Aided Design,
  {ICCAD} 2023, San Francisco, CA, USA, October 28 - Nov. 2, 2023}, pages 1--9.
  {IEEE}, 2023.
\newblock \href {https://doi.org/10.1109/ICCAD57390.2023.10323924}
  {\path{doi:10.1109/ICCAD57390.2023.10323924}}.

\bibitem{DBLP:conf/ecai/ShaikP24}
Irfansha Shaik and Jaco van~de Pol.
\newblock Optimal layout-aware {CNOT} circuit synthesis with qubit permutation.
\newblock In {\em {ECAI} 2024}, volume 392 of {\em Frontiers in Artificial
  Intelligence and Applications}, pages 4207--4215. {IOS} Press, 2024.
\newblock \href {https://doi.org/10.3233/FAIA240993}
  {\path{doi:10.3233/FAIA240993}}.

\bibitem{shaik2024optimal}
Irfansha Shaik and Jaco van~de Pol.
\newblock Optimal layout synthesis for deep quantum circuits on {NISQ}
  processors with 100+ qubits.
\newblock In {\em 27th International Conference on Theory and Applications of
  Satisfiability Testing {SAT}}, volume 305 of {\em LIPIcs}. Schloss Dagstuhl -
  Leibniz-Zentrum f{\"{u}}r Informatik, 2024.

\bibitem{Sivarajah_2021}
Seyon Sivarajah, Silas Dilkes, Alexander Cowtan, Will Simmons, Alec Edgington,
  and Ross Duncan.
\newblock t|ket⟩: a retargetable compiler for nisq devices.
\newblock {\em Quantum Science and Technology}, 6(1):014003, nov 2020.
\newblock URL: \url{https://dx.doi.org/10.1088/2058-9565/ab8e92}, \href
  {https://doi.org/10.1088/2058-9565/ab8e92}
  {\path{doi:10.1088/2058-9565/ab8e92}}.

\bibitem{Wille_2023}
Robert Wille and Lukas Burgholzer.
\newblock {MQT} {QMAP}.
\newblock In {\em Proceedings of ISPD-23}. {ACM}, mar 2023.
\newblock \href {https://doi.org/10.1145/3569052.3578928}
  {\path{doi:10.1145/3569052.3578928}}.

\bibitem{DBLP:conf/dac/WilleBZ19}
Robert Wille, Lukas Burgholzer, and Alwin Zulehner.
\newblock Mapping quantum circuits to {IBM} {QX} architectures using the
  minimal number of {SWAP} and {H} operations.
\newblock In {\em {DAC}-19}, page 142. {ACM}, 2019.
\newblock \href {https://doi.org/10.1145/3316781.3317859}
  {\path{doi:10.1145/3316781.3317859}}.

\end{thebibliography}

\appendix

\section{Redundant Clauses and Auxiliary Variables}
\label{sec:redundantclauses}
In SAT encodings, there is often a trade-off between search and propagation.
In~\cite{shaik2024optimal}, we observed that adding redundant clauses sometimes improve performance of SAT solving.
Giving explicit constraints that trigger unit clause propagation can avoid some unnecessary search.
For example, Equation 4 explicitly encodes that exactly-one CNOT variable can be true at time step $t$.
It also implicitly encodes that exactly-one control or target variable can be true in the same time step.
If a SAT solver sets a CNOT variable to true, then all control, target, and rest of the CNOT variables are propagated.
However, if a SAT solver instead sets a control variable to true it does not learn any more information.
By giving (redundant) exactly-one constraints on control and target variables can avoid this unnecessary search.

Similarly, using auxiliary variables can impact performance of SAT solving.
In Equation 1, recall that propagation variable $\px^{t}_{i,a}$ is true then $\xmatrixvar^{t}_{i,a}$ is propagated
else $\xmatrixvar^{t}_{i,a}$ is flipped.
We experimented with using additional auxiliary variable $\fx^{t}_{i,a}$ to indicate the flipping instead of $\neg\px^{t}_{i,a}$.
We add an option to use the new flipping variables for both $x,z$ state variables as part of XOR constraints in our encoding.
We provide these variations as an option, we turn them on default as they seem to help our backend SAT solver
Cadical~\cite{BiereFazekasFleuryHeisinger-SAT-Competition-2020-solvers}.

\section{Full Tables}

\begin{table}[h]
  \caption{Experiment 1: Random Clifford cnot-count optimization, m: cx-count t: time, TO: timeout. }
  \label{tb:experiment1cxc}
  \centering
  \begin{tabular}{lrrrrrrrrrrrr}
  \toprule
              & \multicolumn{7}{c}{without-permutation} & \multicolumn{5}{c}{with-permutation}\\
              \cmidrule(lr){2-8} \cmidrule(lr){9-13}
  Tool:      & tket & \multicolumn{2}{c}{qmap} & \multicolumn{2}{c}{qsynth-f} &\multicolumn{2}{c}{qsynth-b} & tket & \multicolumn{2}{c}{qsynth-f} &\multicolumn{2}{c}{qsynth-b}\\
  \cmidrule(lr){3-4} \cmidrule(lr){5-6} \cmidrule(lr){7-8} \cmidrule(lr){10-11} \cmidrule(lr){12-13}
  instance & m & m &t & m & t & m & t & m & m & t & m & t \\
  \midrule
  3q05306 & 4 & 4 & 1452.5 & 4 & 0.13 & 4 & 0.06 & 2 & 2 & 0.1 & 2 & 0.09 \\
  3q33936 & 3 & 3 & 29.3 & 3 & 0.07 & 3 & 0.05 & 3 & 3 & 0.07 & 3 & 0.06 \\
  3q50494 & 4 & 4 & 1804.7 & 4 & 0.08 & 4 & 0.05 & 3 & 3 & 0.06 & 3 & 0.05 \\
  3q55125 & 3 & 3 & 77.2 & 3 & 0.06 & 3 & 0.04 & 3 & 3 & 0.06 & 3 & 0.05 \\
  3q99346 & 4 & 4 & 1650.2 & 4 & 0.08 & 4 & 0.05 & 3 & 3 & 0.06 & 3 & 0.05 \\
  4q05306 & 6 & - & TO & 6 & 0.2 & 6 & 0.12 & 6 & 5 & 0.44 & 5 & 0.5 \\
  4q33936 & 8 & - & TO & 6 & 0.38 & 6 & 0.31 & 8 & 5 & 0.35 & 5 & 0.71 \\
  4q50494 & 7 & - & TO & 7 & 0.49 & 7 & 0.38 & 4 & 4 & 0.1 & 4 & 0.07 \\
  4q55125 & 6 & - & TO & 6 & 0.18 & 6 & 0.11 & 6 & 5 & 0.43 & 5 & 0.33 \\
  4q99346 & 7 & - & TO & 5 & 0.19 & 5 & 0.25 & 6 & 5 & 0.43 & 5 & 0.35 \\
  5q05306 & 12 & - & TO & 8 & 4.74 & 8 & 14.1 & 11 & 8 & 34.93 & 8 & 151.2 \\
  5q33936 & 13 & - & TO & 9 & 11.7 & 9 & 33.44 & 10 & 8 & 34.35 & 8 & 126.5 \\
  5q50494 & 17 & - & TO & 8 & 4.03 & 8 & 36.89 & 11 & 7 & 22.35 & 7 & 56.6 \\
  5q55125 & 13 & - & TO & 9 & 13.89 & 9 & 22.23 & 9 & 7 & 4.49 & 7 & 9.25 \\
  5q99346 & 11 & - & TO & 7 & 1.05 & 7 & 2.86 & 10 & 6 & 1.52 & 6 & 7.88 \\
  6q05306 & 15 & - & TO & - & TO & 12 & TO & 15 & 15 & TO & 11 & TO \\
  6q33936 & 20 & - & TO & 11 & 2017.7 & 12 & TO & 14 & 14 & TO & 11 & TO \\
  6q50494 & 17 & - & TO & 11 & 964.3 & 11 & 4525.9 & 14 & 14 & TO & 12 & TO \\
  6q55125 & 18 & - & TO & 11 & 1060.2 & 11 & 5059.7 & 14 & 10 & 6452.9 & 10 & TO \\
  6q99346 & 15 & - & TO & - & TO & 12 & TO & 15 & - & TO & 11 & TO \\
  7q05306 & 25 & - & TO & - & TO & - & TO & 24 & - & TO & 19 & TO \\
  7q33936 & 24 & - & TO & - & TO & - & TO & 20 & - & TO & 17 & TO \\
  7q50494 & 22 & - & TO & - & TO & 18 & TO & 20 & - & TO & - & TO \\
  7q55125 & 22 & - & TO & - & TO & 17 & TO & 18 & - & TO & 17 & TO \\
  7q99346 & 21 & - & TO & - & TO & - & TO & 20 & - & TO & 19 & TO \\
  \bottomrule
  \end{tabular}
  \end{table}

  \begin{table}[t]
    \caption{Experiment 1: Random Clifford cnot-depth optimization, m: cx-depth, t: time, TO: timeout. }
    \label{tb:experiment1cxd}
    \centering
    \begin{tabular}{lrrrrrrrrrrrr}
    \toprule
    & \multicolumn{7}{c}{without-permutation} & \multicolumn{5}{c}{with-permutation}\\
    \cmidrule(lr){2-8} \cmidrule(lr){9-13}
    Tool:      & tket & \multicolumn{2}{c}{qmap} & \multicolumn{2}{c}{qsynth-f} &\multicolumn{2}{c}{qsynth-b} & tket & \multicolumn{2}{c}{qsynth-f} &\multicolumn{2}{c}{qsynth-b}\\
    \cmidrule(lr){3-4} \cmidrule(lr){5-6} \cmidrule(lr){7-8} \cmidrule(lr){10-11} \cmidrule(lr){12-13}
    instance & m & m &t & m & t & m & t & m & m & t & m & t \\
    \midrule
    3q05306 & 4 & 4 & 0.4 & 4 & 0.13 & 4 & 0.06 & 2 & 2 & 0.1 & 2 & 0.09 \\
    3q33936 & 3 & 3 & 0.39 & 3 & 0.07 & 3 & 0.05 & 3 & 3 & 0.07 & 3 & 0.05 \\
    3q50494 & 4 & 4 & 0.25 & 4 & 0.08 & 4 & 0.05 & 3 & 3 & 0.06 & 3 & 0.04 \\
    3q55125 & 3 & 3 & 0.42 & 3 & 0.06 & 3 & 0.04 & 3 & 3 & 0.06 & 3 & 0.05 \\
    3q99346 & 4 & 4 & 0.34 & 4 & 0.08 & 4 & 0.05 & 3 & 3 & 0.06 & 3 & 0.05 \\
    4q05306 & 4 & 5 & 4.0 & 4 & 0.11 & 4 & 0.07 & 4 & 3 & 0.12 & 3 & 0.1 \\
    4q33936 & 7 & 5 & 4.27 & 4 & 0.16 & 4 & 0.34 & 7 & 3 & 0.14 & 3 & 0.57 \\
    4q50494 & 6 & 6 & 6.85 & 4 & 0.17 & 4 & 0.17 & 3 & 3 & 0.07 & 3 & 0.05 \\
    4q55125 & 4 & 4 & 2.53 & 3 & 0.1 & 3 & 0.09 & 4 & 3 & 0.12 & 3 & 0.1 \\
    4q99346 & 7 & 4 & 8.08 & 3 & 0.11 & 3 & 0.44 & 6 & 3 & 0.21 & 3 & 0.36 \\
    5q05306 & 10 & 8 & 437.1 & 5 & 1.64 & 5 & 9.68 & 8 & 4 & 2.57 & 4 & 4.56 \\
    5q33936 & 10 & 7 & 533.8 & 5 & 0.91 & 5 & 6.3 & 8 & 4 & 2.81 & 4 & 7.0 \\
    5q50494 & 15 & 7 & 626.2 & 5 & 2.25 & 5 & 40.49 & 9 & 4 & 5.71 & 4 & 6.98 \\
    5q55125 & 12 & 8 & 347.7 & 5 & 0.91 & 5 & 21.45 & 8 & 4 & 1.27 & 4 & 4.98 \\
    5q99346 & 9 & 6 & 226.7 & 4 & 0.39 & 4 & 3.25 & 8 & 4 & 0.37 & 4 & 5.41 \\
    6q05306 & 11 & - & TO & 5 & 23.59 & 5 & 299.27 & 11 & 5 & 168.09 & 5 & 607.01 \\
    6q33936 & 16 & 11 & TO & 5 & 6.39 & 5 & 994.98 & 10 & 5 & 60.75 & 5 & 379.15 \\
    6q50494 & 14 & 14 & TO & 5 & 1.86 & 5 & 741.38 & 11 & 5 & 64.86 & 5 & 357.08 \\
    6q55125 & 15 & 17 & TO & 5 & 3.03 & 5 & 638.7 & 12 & 5 & 35.44 & 5 & 587.76 \\
    6q99346 & 12 & 11 & TO & 5 & 17.66 & 5 & 1522.43 & 12 & 5 & 82.35 & 5 & 984.03 \\
    7q05306 & 19 & - & TO & - & TO & 18 & TO & 18 & - & TO & - & TO \\
    7q33936 & 19 & - & TO & 6 & 4151.7 & - & TO & 16 & - & TO & - & TO \\
    7q50494 & 18 & - & TO & 6 & 7347.0 & - & TO & 15 & - & TO & - & TO \\
    7q55125 & 19 & - & TO & 6 & 1485.5 & - & TO & 14 & - & TO & 12 & TO \\
    7q99346 & 15 & - & TO & 6 & 241.0 & - & TO & 14 & - & TO & - & TO \\
    \bottomrule
    \end{tabular}
    \end{table}

\begin{table}[htbp]
  \caption{Experiment 2a: VQE cnot-count optimization, m: cx-count, t: time. }
  \label{tb:experiment2a_vqe_cxc}
  \centering
  \rotatebox{90}{
  \begin{tabular}{llllllllllll}
  \toprule
  & & \multicolumn{5}{c}{without-permutation} & \multicolumn{5}{c}{with-permutation}\\
  \cmidrule(lr){3-7} \cmidrule(lr){8-12}
  Tools: & & tket & \multicolumn{2}{c}{qsynth-f} & \multicolumn{2}{c}{tket+qsynth-f} & tket & \multicolumn{2}{c}{qsynth} & \multicolumn{2}{c}{tket+qsynth-f}\\
  \cmidrule(lr){4-5} \cmidrule(lr){6-7} \cmidrule(lr){9-10} \cmidrule(lr){11-12}
  instance & org & m & m (ch\%) & t & m (ch\%) & t & m (ch\%) & m (ch\%) & t & m (ch\%) & t \\
  \midrule
  pennylane\_8q0 & 316 & 292 (-7.6) & 291 (-7.9) & 34.2 & 289 (-8.5) & 20.9 & 289 (-8.5) & 288 (-8.9) & 130.5 & 285 (-9.8) & 71.8 \\
  pennylane\_8q1 & 376 & 347 (-7.7) & 345 (-8.2) & 149.5 & 345 (-8.2) & 127.6 & 346 (-8.0) & 344 (-8.5) & 601.1 & 343 (-8.8) & 601.2 \\
  pennylane\_16q0 & 800 & 746 (-6.8) & 765 (-4.4) & 602.6 & 745 (-6.9) & 601.9 & 746 (-6.8) & 777 (-2.9) & 602.0 & 745 (-6.9) & 601.2 \\
  pennylane\_16q1 & 860 & 801 (-6.9) & 818 (-4.9) & 603.1 & 795 (-7.6) & 601.5 & 799 (-7.1) & 821 (-4.5) & 602.6 & 793 (-7.8) & 601.6 \\
  simple\_8q0 & 316 & 310 (-1.9) & 311 (-1.6) & 17.0 & 309 (-2.2) & 17.6 & 290 (-8.2) & 292 (-7.6) & 40.3 & 289 (-8.5) & 17.4 \\
  simple\_8q1 & 376 & 367 (-2.4) & 364 (-3.2) & 22.0 & 364 (-3.2) & 21.0 & 344 (-8.5) & 350 (-6.9) & 34.1 & 343 (-8.8) & 31.3 \\
  simple\_16q0 & 800 & 788 (-1.5) & 791 (-1.1) & 603.8 & 788 (-1.5) & 602.1 & 743 (-7.1) & 776 (-3.0) & 603.2 & 743 (-7.1) & 601.6 \\
  simple\_16q1 & 860 & 855 (-0.6) & 855 (-0.6) & 603.9 & 855 (-0.6) & 603.5 & 800 (-7.0) & 837 (-2.7) & 602.7 & 799 (-7.1) & 603.3 \\
  yardanov\_8q0 & 298 & 292 (-2.0) & 293 (-1.7) & 8.3 & 291 (-2.3) & 8.2 & 290 (-2.7) & 291 (-2.3) & 16.4 & 289 (-3.0) & 16.2 \\
  yardanov\_8q1 & 353 & 344 (-2.5) & 341 (-3.4) & 10.6 & 341 (-3.4) & 10.0 & 343 (-2.8) & 340 (-3.7) & 18.5 & 340 (-3.7) & 17.4 \\
  yardanov\_16q0 & 750 & 738 (-1.6) & 739 (-1.5) & 603.9 & 737 (-1.7) & 602.6 & 738 (-1.6) & 741 (-1.2) & 602.7 & 738 (-1.6) & 601.8 \\
  yardanov\_16q1 & 805 & 800 (-0.6) & 799 (-0.7) & 604.0 & 799 (-0.7) & 604.4 & 799 (-0.7) & 799 (-0.7) & 602.8 & 799 (-0.7) & 603.2 \\
  \bottomrule
  \end{tabular}
  }
  \end{table}
  
  \begin{table}[htbp]
  \caption{Experiment 2a: VQE cnot-depth optimization, m: cx-depth, t: time. }
  \label{tb:experiment2a_vqe_cxd}
  \centering
  \rotatebox{90}{
  \begin{tabular}{llllllllllll}
  \toprule
  & & \multicolumn{5}{c}{without-permutation} & \multicolumn{5}{c}{with-permutation}\\
  \cmidrule(lr){3-7} \cmidrule(lr){8-12}
  Tools: & & tket & \multicolumn{2}{c}{qsynth-f} & \multicolumn{2}{c}{tket+qsynth-f} & tket & \multicolumn{2}{c}{qsynth} & \multicolumn{2}{c}{tket+qsynth-f}\\
  \cmidrule(lr){4-5} \cmidrule(lr){6-7} \cmidrule(lr){9-10} \cmidrule(lr){11-12}
  instance & org & m (ch\%)  & m (ch\%)  & t & m (ch\%)  & t & m (ch\%)  & m (ch\%)  & t & m (ch\%)  & t \\
  \midrule
  pennylane\_8q0 & 233 & 214 (-8.2) & 199 (-14.6) & 8.6 & 197 (-15.5) & 9.1 & 213 (-8.6) & 198 (-15.0) & 18.1 & 196 (-15.9) & 25.2 \\
  pennylane\_8q1 & 275 & 256 (-6.9) & 231 (-16.0) & 12.4 & 229 (-16.7) & 11.5 & 255 (-7.3) & 229 (-16.7) & 36.0 & 228 (-17.1) & 27.2 \\
  pennylane\_16q0 & 428 & 395 (-7.7) & 372 (-13.1) & 45.8 & 366 (-14.5) & 41.5 & 395 (-7.7) & 374 (-12.6) & 197.3 & 363 (-15.2) & 200.2 \\
  pennylane\_16q1 & 479 & 444 (-7.3) & 414 (-13.6) & 60.5 & 406 (-15.2) & 51.9 & 443 (-7.5) & 417 (-12.9) & 363.5 & 407 (-15.0) & 230.6 \\
  simple\_8q0 & 252 & 250 (-0.8) & 241 (-4.4) & 5.4 & 238 (-5.6) & 5.7 & 241 (-4.4) & 220 (-12.7) & 7.5 & 221 (-12.3) & 8.2 \\
  simple\_8q1 & 300 & 300 (0.0) & 276 (-8.0) & 7.8 & 277 (-7.7) & 7.6 & 292 (-2.7) & 266 (-11.3) & 14.4 & 268 (-10.7) & 14.8 \\
  simple\_16q0 & 470 & 465 (-1.1) & 435 (-7.4) & 29.6 & 437 (-7.0) & 31.1 & 453 (-3.6) & 418 (-11.1) & 65.6 & 419 (-10.9) & 68.0 \\
  simple\_16q1 & 536 & 534 (-0.4) & 494 (-7.8) & 34.3 & 495 (-7.6) & 33.7 & 513 (-4.3) & 475 (-11.4) & 105.4 & 481 (-10.3) & 72.4 \\
  yardanov\_8q0 & 234 & 232 (-0.9) & 227 (-3.0) & 3.1 & 225 (-3.8) & 3.4 & 232 (-0.9) & 225 (-3.8) & 4.9 & 223 (-4.7) & 4.8 \\
  yardanov\_8q1 & 277 & 277 (0.0) & 259 (-6.5) & 3.6 & 259 (-6.5) & 3.8 & 277 (0.0) & 260 (-6.1) & 6.1 & 259 (-6.5) & 6.0 \\
  yardanov\_16q0 & 434 & 429 (-1.2) & 400 (-7.8) & 13.7 & 401 (-7.6) & 14.4 & 429 (-1.2) & 401 (-7.6) & 21.2 & 401 (-7.6) & 21.8 \\
  yardanov\_16q1 & 493 & 491 (-0.4) & 457 (-7.3) & 16.2 & 457 (-7.3) & 16.3 & 491 (-0.4) & 454 (-7.9) & 26.3 & 452 (-8.3) & 26.4 \\
  \bottomrule
  \end{tabular}
  }
  \end{table}

\begin{table}[htbp]
  \caption{Experiment 2b: Feynman cnot-count optimization, m: cx-count, q: qubits, t: time. }
  \label{tb:experiment2b_feynman_cxc}
  \centering
  \resizebox{\columnwidth}{!}{%
  \rotatebox{90}{
  \begin{tabular}{lllllllllllll}
  \toprule
  & & & \multicolumn{5}{c}{without-permutation} & \multicolumn{5}{c}{with-permutation}\\
  \cmidrule(lr){4-8} \cmidrule(lr){9-13}
  Tools: & & & tket & \multicolumn{2}{c}{qsynth-f} & \multicolumn{2}{c}{tket+qsynth-f} & tket & \multicolumn{2}{c}{qsynth} & \multicolumn{2}{c}{tket+qsynth-f}\\
  \cmidrule(lr){5-6} \cmidrule(lr){7-8} \cmidrule(lr){10-11} \cmidrule(lr){12-13}
  instance & q & org & m (ch\%)  & m (ch\%)  & t & m (ch\%)  & t & m (ch\%)  & m (ch\%)  & t & m (ch\%)  & t \\
  \midrule
  tof\_3 & 5 & 18 & 18 (0.0) & 18 (0.0) & 0.2 & 18 (0.0) & 0.2 & 18 (0.0) & 18 (0.0) & 0.2 & 18 (0.0) & 0.2 \\
  barenco\_tof\_3 & 5 & 24 & 24 (0.0) & 23 (-4.2) & 0.3 & 23 (-4.2) & 0.3 & 24 (0.0) & 23 (-4.2) & 0.3 & 23 (-4.2) & 0.3 \\
  mod5\_4 & 5 & 28 & 20 (-28.6) & 19 (-32.1) & 0.3 & 19 (-32.1) & 0.2 & 20 (-28.6) & 19 (-32.1) & 0.4 & 19 (-32.1) & 0.2 \\
  qft\_4 & 5 & 46 & 46 (0.0) & 45 (-2.2) & 0.9 & 45 (-2.2) & 0.9 & 46 (0.0) & 45 (-2.2) & 1.4 & 45 (-2.2) & 1.5 \\
  tof\_4 & 7 & 30 & 30 (0.0) & 29 (-3.3) & 0.4 & 29 (-3.3) & 0.5 & 30 (0.0) & 29 (-3.3) & 0.5 & 29 (-3.3) & 0.5 \\
  barenco\_tof\_4 & 7 & 48 & 42 (-12.5) & 39 (-18.8) & 1.2 & 39 (-18.8) & 0.6 & 42 (-12.5) & 39 (-18.8) & 0.9 & 39 (-18.8) & 0.7 \\
  hwb6 & 7 & 116 & 112 (-3.4) & 108 (-6.9) & 4.1 & 107 (-7.8) & 3.4 & 106 (-8.6) & 105 (-9.5) & 7.5 & 103 (-11.2) & 3.1 \\
  tof\_5 & 9 & 42 & 42 (0.0) & 40 (-4.8) & 0.9 & 40 (-4.8) & 1.0 & 42 (0.0) & 40 (-4.8) & 1.0 & 40 (-4.8) & 1.2 \\
  mod\_mult\_55 & 9 & 48 & 48 (0.0) & 46 (-4.2) & 3.3 & 46 (-4.2) & 3.4 & 48 (0.0) & 46 (-4.2) & 6.7 & 46 (-4.2) & 7.1 \\
  barenco\_tof\_5 & 9 & 72 & 60 (-16.7) & 55 (-23.6) & 1.3 & 55 (-23.6) & 1.1 & 60 (-16.7) & 55 (-23.6) & 1.6 & 55 (-23.6) & 1.2 \\
  grover\_5 & 9 & 288 & 206 (-28.5) & 219 (-24.0) & 4.5 & 200 (-30.6) & 3.3 & 206 (-28.5) & 219 (-24.0) & 5.3 & 200 (-30.6) & 3.3 \\
  vbe\_adder\_3 & 10 & 70 & 56 (-20.0) & 66 (-5.7) & 601.8 & 56 (-20.0) & 601.9 & 54 (-22.9) & 65 (-7.1) & 601.5 & 52 (-25.7) & 252.1 \\
  fprenor & 10 & 121 & 97 (-19.8) & 97 (-19.8) & 8.0 & 95 (-21.5) & 7.7 & 94 (-22.3) & 97 (-19.8) & 38.8 & 92 (-24.0) & 27.5 \\
  mod\_red\_21 & 11 & 105 & 98 (-6.7) & 100 (-4.8) & 2.2 & 96 (-8.6) & 2.1 & 98 (-6.7) & 100 (-4.8) & 2.8 & 96 (-8.6) & 2.6 \\
  gf2\^4\_mult & 12 & 99 & 99 (0.0) & 99 (0.0) & 2.8 & 99 (0.0) & 2.8 & 99 (0.0) & 99 (0.0) & 3.7 & 99 (0.0) & 3.7 \\
  rc\_adder\_6 & 14 & 93 & 91 (-2.2) & 91 (-2.2) & 604.6 & 91 (-2.2) & 604.7 & 86 (-7.5) & 86 (-7.5) & 603.2 & 86 (-7.5) & 603.8 \\
  csla\_mux\_3 & 15 & 80 & 80 (0.0) & 80 (0.0) & 36.5 & 80 (0.0) & 40.6 & 75 (-6.2) & 75 (-6.2) & 47.8 & 75 (-6.2) & 17.2 \\
  gf2\^5\_mult & 15 & 154 & 154 (0.0) & 154 (0.0) & 9.7 & 154 (0.0) & 9.6 & 154 (0.0) & 154 (0.0) & 20.9 & 154 (0.0) & 19.4 \\
  ham15-low & 17 & 236 & 223 (-5.5) & 227 (-3.8) & 603.6 & 218 (-7.6) & 603.7 & 212 (-10.2) & 228 (-3.4) & 603.1 & 207 (-12.3) & 602.9 \\
  ham15-med & 17 & 534 & 407 (-23.8) & 425 (-20.4) & 607.6 & 371 (-30.5) & 606.9 & 396 (-25.8) & 425 (-20.4) & 605.2 & 360 (-32.6) & 604.8 \\
  gf2\^6\_mult & 18 & 221 & 221 (0.0) & 221 (0.0) & 51.3 & 221 (0.0) & 51.5 & 221 (0.0) & 221 (0.0) & 335.4 & 221 (0.0) & 321.7 \\
  tof\_10 & 19 & 102 & 102 (0.0) & 95 (-6.9) & 610.1 & 95 (-6.9) & 610.4 & 102 (0.0) & 95 (-6.9) & 608.2 & 95 (-6.9) & 608.2 \\
  barenco\_tof\_10 & 19 & 192 & 150 (-21.9) & 135 (-29.7) & 3.8 & 135 (-29.7) & 3.7 & 150 (-21.9) & 135 (-29.7) & 5.3 & 135 (-29.7) & 3.9 \\
  ham15-high & 20 & 2149 & 1630 (-24.2) & 2033 (-5.4) & 606.8 & 1594 (-25.8) & 606.9 & 1620 (-24.6) & 2032 (-5.4) & 606.3 & 1584 (-26.3) & 605.9 \\
  gf2\^7\_mult & 21 & 300 & 300 (0.0) & 300 (0.0) & 208.1 & 300 (0.0) & 247.4 & 300 (0.0) & 300 (0.0) & 602.1 & 300 (0.0) & 601.9 \\
  qcla\_com\_7 & 24 & 186 & 158 (-15.1) & 176 (-5.4) & 611.6 & 158 (-15.1) & 613.3 & 154 (-17.2) & 180 (-3.2) & 608.7 & 154 (-17.2) & 608.8 \\
  gf2\^8\_mult & 24 & 405 & 405 (0.0) & 405 (0.0) & 602.9 & 405 (0.0) & 603.0 & 402 (-0.7) & 405 (0.0) & 602.6 & 402 (-0.7) & 602.4 \\
  adder\_8 & 24 & 409 & 353 (-13.7) & 377 (-7.8) & 612.3 & 350 (-14.4) & 613.5 & 353 (-13.7) & 383 (-6.4) & 610.3 & 350 (-14.4) & 609.5 \\
  \bottomrule
  \end{tabular}
  }}
  \end{table}
  
  \begin{table}[htbp]
  \caption{Experiment 2b: Feynman cnot-depth optimization, m: cx-depth, q: qubits, t: time. }
  \label{tb:experiment2b_feynman_cxd}
  \centering
  \resizebox{\columnwidth}{!}{%
  \rotatebox{90}{
  \begin{tabular}{lllllllllllll}
  \toprule
  & & & \multicolumn{5}{c}{without-permutation} & \multicolumn{5}{c}{with-permutation}\\
  \cmidrule(lr){4-8} \cmidrule(lr){9-13}
  Tools: & & & tket & \multicolumn{2}{c}{qsynth-f} & \multicolumn{2}{c}{tket+qsynth-f} & tket & \multicolumn{2}{c}{qsynth} & \multicolumn{2}{c}{tket+qsynth-f}\\
  \cmidrule(lr){5-6} \cmidrule(lr){7-8} \cmidrule(lr){10-11} \cmidrule(lr){12-13}
  instance & q & org & m (ch\%)  & m (ch\%)  & t & m (ch\%)  & t & m (ch\%)  & m (ch\%)  & t & m (ch\%)  & t \\
  \midrule
  tof\_3 & 5 & 16 & 16 (0.0) & 16 (0.0) & 0.2 & 16 (0.0) & 0.2 & 16 (0.0) & 16 (0.0) & 0.2 & 16 (0.0) & 0.2 \\
  barenco\_tof\_3 & 5 & 22 & 22 (0.0) & 21 (-4.5) & 0.3 & 21 (-4.5) & 0.2 & 22 (0.0) & 21 (-4.5) & 0.3 & 21 (-4.5) & 0.3 \\
  mod5\_4 & 5 & 27 & 19 (-29.6) & 14 (-48.1) & 0.2 & 14 (-48.1) & 0.2 & 19 (-29.6) & 14 (-48.1) & 0.2 & 14 (-48.1) & 0.2 \\
  qft\_4 & 5 & 43 & 43 (0.0) & 39 (-9.3) & 0.7 & 39 (-9.3) & 0.7 & 43 (0.0) & 39 (-9.3) & 0.9 & 39 (-9.3) & 1.0 \\
  tof\_4 & 7 & 26 & 26 (0.0) & 25 (-3.8) & 0.4 & 25 (-3.8) & 0.4 & 26 (0.0) & 25 (-3.8) & 0.4 & 25 (-3.8) & 0.4 \\
  barenco\_tof\_4 & 7 & 42 & 39 (-7.1) & 34 (-19.0) & 0.7 & 34 (-19.0) & 0.6 & 39 (-7.1) & 34 (-19.0) & 0.9 & 34 (-19.0) & 0.7 \\
  hwb6 & 7 & 96 & 92 (-4.2) & 82 (-14.6) & 1.8 & 81 (-15.6) & 1.9 & 86 (-10.4) & 80 (-16.7) & 2.4 & 79 (-17.7) & 2.4 \\
  mod\_mult\_55 & 9 & 29 & 29 (0.0) & 27 (-6.9) & 0.7 & 27 (-6.9) & 0.9 & 29 (0.0) & 27 (-6.9) & 1.2 & 27 (-6.9) & 1.3 \\
  tof\_5 & 9 & 36 & 36 (0.0) & 34 (-5.6) & 0.6 & 34 (-5.6) & 0.6 & 36 (0.0) & 34 (-5.6) & 0.7 & 34 (-5.6) & 0.6 \\
  barenco\_tof\_5 & 9 & 62 & 56 (-9.7) & 46 (-25.8) & 1.3 & 46 (-25.8) & 1.1 & 56 (-9.7) & 46 (-25.8) & 1.7 & 46 (-25.8) & 1.6 \\
  grover\_5 & 9 & 248 & 178 (-28.2) & 180 (-27.4) & 3.5 & 174 (-29.8) & 3.3 & 178 (-28.2) & 180 (-27.4) & 4.4 & 174 (-29.8) & 2.9 \\
  vbe\_adder\_3 & 10 & 52 & 37 (-28.8) & 38 (-26.9) & 2.9 & 32 (-38.5) & 1.0 & 38 (-26.9) & 36 (-30.8) & 1.8 & 31 (-40.4) & 1.4 \\
  fprenor & 10 & 94 & 78 (-17.0) & 70 (-25.5) & 2.1 & 69 (-26.6) & 1.9 & 75 (-20.2) & 69 (-26.6) & 4.4 & 66 (-29.8) & 3.6 \\
  mod\_red\_21 & 11 & 91 & 84 (-7.7) & 83 (-8.8) & 1.8 & 80 (-12.1) & 1.6 & 84 (-7.7) & 83 (-8.8) & 3.3 & 80 (-12.1) & 2.7 \\
  gf2\^4\_mult & 12 & 65 & 65 (0.0) & 64 (-1.5) & 1.3 & 64 (-1.5) & 1.3 & 65 (0.0) & 64 (-1.5) & 1.4 & 64 (-1.5) & 1.5 \\
  rc\_adder\_6 & 14 & 65 & 65 (0.0) & 62 (-4.6) & 19.5 & 62 (-4.6) & 19.4 & 64 (-1.5) & 61 (-6.2) & 337.3 & 61 (-6.2) & 335.0 \\
  csla\_mux\_3 & 15 & 41 & 41 (0.0) & 41 (0.0) & 1.3 & 41 (0.0) & 1.2 & 41 (0.0) & 43 (+4.9) & 1.8 & 41 (0.0) & 1.0 \\
  gf2\^5\_mult & 15 & 86 & 86 (0.0) & 85 (-1.2) & 2.4 & 84 (-2.3) & 2.3 & 86 (0.0) & 85 (-1.2) & 2.7 & 85 (-1.2) & 2.8 \\
  ham15-low & 17 & 183 & 171 (-6.6) & 163 (-10.9) & 601.7 & 157 (-14.2) & 601.8 & 161 (-12.0) & 163 (-10.9) & 601.5 & 150 (-18.0) & 601.3 \\
  ham15-med & 17 & 478 & 367 (-23.2) & 328 (-31.4) & 602.6 & 296 (-38.1) & 602.7 & 359 (-24.9) & 327 (-31.6) & 602.1 & 287 (-40.0) & 602.1 \\
  gf2\^6\_mult & 18 & 107 & 107 (0.0) & 109 (+1.9) & 5.0 & 107 (0.0) & 5.8 & 107 (0.0) & 109 (+1.9) & 8.8 & 109 (+1.9) & 8.2 \\
  tof\_10 & 19 & 86 & 86 (0.0) & 79 (-8.1) & 1.7 & 79 (-8.1) & 1.9 & 86 (0.0) & 79 (-8.1) & 2.0 & 79 (-8.1) & 1.9 \\
  barenco\_tof\_10 & 19 & 162 & 141 (-13.0) & 106 (-34.6) & 3.7 & 106 (-34.6) & 3.3 & 141 (-13.0) & 106 (-34.6) & 5.8 & 106 (-34.6) & 4.9 \\
  ham15-high & 20 & 1878 & 1385 (-26.3) & 1642 (-12.6) & 604.3 & 1287 (-31.5) & 604.2 & 1376 (-26.7) & 1634 (-13.0) & 603.8 & 1276 (-32.1) & 603.7 \\
  gf2\^7\_mult & 21 & 128 & 128 (0.0) & 130 (+1.6) & 7.6 & 128 (0.0) & 7.5 & 128 (0.0) & 130 (+1.6) & 16.2 & 130 (+1.6) & 18.8 \\
  qcla\_com\_7 & 24 & 51 & 47 (-7.8) & 49 (-3.9) & 17.2 & 48 (-5.9) & 13.9 & 48 (-5.9) & 47 (-7.8) & 64.1 & 47 (-7.8) & 67.3 \\
  adder\_8 & 24 & 138 & 127 (-8.0) & 118 (-14.5) & 24.7 & 114 (-17.4) & 23.6 & 127 (-8.0) & 117 (-15.2) & 55.4 & 118 (-14.5) & 47.9 \\
  gf2\^8\_mult & 24 & 155 & 155 (0.0) & 157 (+1.3) & 601.2 & 155 (0.0) & 601.2 & 154 (-0.6) & 157 (+1.3) & 601.0 & 156 (+0.6) & 601.0 \\
  \bottomrule
  \end{tabular}
  }}
  \end{table}

\begin{table}[htbp]
  \caption{Experiment 3a: Mapped VQE cnot-count optimization, m: cx-count, q: qubits, t: time. }
  \label{tb:experiment3a_vqe_cxc}
  \centering
  \rotatebox{90}{
  \begin{tabular}{llllllllll}
  \toprule
  & \multicolumn{3}{c}{sycamore-54}& \multicolumn{3}{c}{rigetti-80}& \multicolumn{3}{c}{eagle-127}\\
  \cmidrule(lr){2-4} \cmidrule(lr){5-7} \cmidrule(lr){8-10}
  & org & \multicolumn{2}{c}{qsynth-f}& org & \multicolumn{2}{c}{qsynth-f}& org & \multicolumn{2}{c}{qsynth-f}\\
  \cmidrule(lr){3-4} \cmidrule(lr){6-7} \cmidrule(lr){9-10}
  instance & m  & m (ch\%)  & t & m  & m (ch\%)  & t & m  & m (ch\%)  & t \\
  \midrule
  pennylane\_8q0 & 646 & 563 (-12.8) & 603.0 & 763 & 616 (-19.3) & 602.7 & 910 & 767 (-15.7) & 604.5 \\
  pennylane\_8q1 & 766 & 644 (-15.9) & 602.2 & 919 & 752 (-18.2) & 603.8 & 1078 & 881 (-18.3) & 605.2 \\
  pennylane\_16q0 & 2024 & 1916 (-5.3) & 603.0 & 2456 & 2310 (-5.9) & 605.1 & 2960 & 2805 (-5.2) & 604.8 \\
  pennylane\_16q1 & 2179 & 2005 (-8.0) & 604.4 & 2698 & 2499 (-7.4) & 603.7 & 3172 & 2991 (-5.7) & 606.2 \\
  simple\_8q0 & 542 & 490 (-9.6) & 602.9 & 662 & 593 (-10.4) & 602.9 & 770 & 666 (-13.5) & 604.9 \\
  simple\_8q1 & 644 & 579 (-10.1) & 603.0 & 722 & 657 (-9.0) & 604.0 & 947 & 814 (-14.0) & 603.9 \\
  simple\_16q0 & 1706 & 1672 (-2.0) & 607.4 & 2171 & 2110 (-2.8) & 603.3 & 2642 & 2564 (-3.0) & 604.5 \\
  simple\_16q1 & 1844 & 1781 (-3.4) & 605.9 & 2354 & 2299 (-2.3) & 604.3 & 2813 & 2739 (-2.6) & 607.2 \\
  yardanov\_8q0 & 515 & 467 (-9.3) & 602.8 & 644 & 569 (-11.6) & 603.5 & 731 & 643 (-12.0) & 604.7 \\
  yardanov\_8q1 & 634 & 579 (-8.7) & 603.6 & 778 & 690 (-11.3) & 604.1 & 871 & 720 (-17.3) & 158.1 \\
  yardanov\_16q0 & 1710 & 1644 (-3.9) & 604.9 & 2181 & 2117 (-2.9) & 604.3 & 2532 & 2468 (-2.5) & 605.3 \\
  yardanov\_16q1 & 1834 & 1775 (-3.2) & 606.4 & 2254 & 2160 (-4.2) & 603.6 & 2689 & 2593 (-3.6) & 606.2 \\
  \bottomrule
  \end{tabular}
  }
  \end{table}
  
  \begin{table}[htbp]
  \caption{Experiment 3a: Mapped VQE cnot-depth optimization, m: cx-depth, q: qubits, t: time. }
  \label{tb:experiment3a_vqe_cxd}
  \centering
  \rotatebox{90}{
  \begin{tabular}{llllllllll}
  \toprule
  & \multicolumn{3}{c}{sycamore-54}& \multicolumn{3}{c}{rigetti-80}& \multicolumn{3}{c}{eagle-127}\\
  \cmidrule(lr){2-4} \cmidrule(lr){5-7} \cmidrule(lr){8-10}
  & org & \multicolumn{2}{c}{qsynth-f}& org & \multicolumn{2}{c}{qsynth-f}& org & \multicolumn{2}{c}{qsynth-f}\\
  \cmidrule(lr){3-4} \cmidrule(lr){6-7} \cmidrule(lr){9-10}
  instance & m  & m (ch\%)  & t & m  & m (ch\%)  & t & m  & m (ch\%)  & t \\
  \midrule
  pennylane\_8q0 & 525 & 405 (-22.9) & 35.8 & 600 & 453 (-24.5) & 35.9 & 724 & 545 (-24.7) & 67.4 \\
  pennylane\_8q1 & 610 & 443 (-27.4) & 31.7 & 717 & 531 (-25.9) & 32.5 & 864 & 650 (-24.8) & 151.3 \\
  pennylane\_16q0 & 1159 & 973 (-16.0) & 415.2 & 1419 & 1160 (-18.3) & 605.3 & 1569 & 1330 (-15.2) & 607.9 \\
  pennylane\_16q1 & 1377 & 1138 (-17.4) & 607.8 & 1579 & 1363 (-13.7) & 605.1 & 1886 & 1508 (-20.0) & 611.9 \\
  simple\_8q0 & 442 & 362 (-18.1) & 24.0 & 534 & 452 (-15.4) & 35.2 & 632 & 519 (-17.9) & 104.7 \\
  simple\_8q1 & 526 & 428 (-18.6) & 33.6 & 609 & 479 (-21.3) & 449.6 & 781 & 596 (-23.7) & 86.5 \\
  simple\_16q0 & 1098 & 995 (-9.4) & 605.3 & 1286 & 1099 (-14.5) & 606.8 & 1536 & 1287 (-16.2) & 608.7 \\
  simple\_16q1 & 1114 & 990 (-11.1) & 604.0 & 1393 & 1192 (-14.4) & 606.7 & 1725 & 1439 (-16.6) & 608.6 \\
  yardanov\_8q0 & 419 & 356 (-15.0) & 38.6 & 514 & 425 (-17.3) & 31.0 & 597 & 480 (-19.6) & 33.0 \\
  yardanov\_8q1 & 474 & 415 (-12.4) & 61.9 & 636 & 518 (-18.6) & 28.5 & 682 & 549 (-19.5) & 36.7 \\
  yardanov\_16q0 & 1017 & 897 (-11.8) & 272.4 & 1209 & 1096 (-9.3) & 606.4 & 1358 & 1191 (-12.3) & 607.9 \\
  yardanov\_16q1 & 1056 & 960 (-9.1) & 604.6 & 1280 & 1117 (-12.7) & 607.7 & 1451 & 1218 (-16.1) & 609.1 \\
  \bottomrule
  \end{tabular}
  }
  \end{table}

\begin{table}[htbp]
  \caption{Experiment 3b: Mapped Feynman cnot-count optimization, m: cx-count, q: qubits, t: time. }
  \label{tb:experiment3b_feynman_cxc}
  \centering
  \resizebox{\columnwidth}{!}{%
  \rotatebox{90}{
  \begin{tabular}{lllllllllll}
  \toprule
  & & \multicolumn{3}{c}{sycamore-54}& \multicolumn{3}{c}{rigetti-80}& \multicolumn{3}{c}{eagle-127}\\
  \cmidrule(lr){3-5} \cmidrule(lr){6-8} \cmidrule(lr){9-11}
  & &org & \multicolumn{2}{c}{qsynth-f}& org & \multicolumn{2}{c}{qsynth-f}& org & \multicolumn{2}{c}{qsynth-f}\\
  \cmidrule(lr){4-5} \cmidrule(lr){7-8} \cmidrule(lr){10-11}
  instance & q & m  & m (ch\%)  & t & m  & m (ch\%)  & t & m  & m (ch\%)  & t \\
  \midrule
  barenco\_tof\_3 & 5 & 48 & 40 (-16.7) & 0.9 & 48 & 38 (-20.8) & 0.8 & 51 & 37 (-27.5) & 0.8 \\
  tof\_3 & 5 & 33 & 30 (-9.1) & 0.7 & 39 & 30 (-23.1) & 0.5 & 39 & 30 (-23.1) & 0.6 \\
  qft\_4 & 5 & 100 & 79 (-21.0) & 2.3 & 106 & 79 (-25.5) & 1.6 & 109 & 85 (-22.0) & 1.9 \\
  mod5\_4 & 5 & 35 & 31 (-11.4) & 1.1 & 38 & 34 (-10.5) & 1.5 & 47 & 33 (-29.8) & 0.8 \\
  hwb6 & 7 & 229 & 183 (-20.1) & 36.3 & 253 & 202 (-20.2) & 25.7 & 274 & 222 (-19.0) & 13.9 \\
  barenco\_tof\_4 & 7 & 78 & 66 (-15.4) & 4.2 & 78 & 64 (-17.9) & 1.7 & 84 & 67 (-20.2) & 1.4 \\
  tof\_4 & 7 & 63 & 52 (-17.5) & 1.6 & 66 & 48 (-27.3) & 1.4 & 72 & 56 (-22.2) & 1.4 \\
  barenco\_tof\_5 & 9 & 111 & 94 (-15.3) & 4.9 & 117 & 94 (-19.7) & 2.1 & 135 & 102 (-24.4) & 2.7 \\
  mod\_mult\_55 & 9 & 105 & 79 (-24.8) & 5.6 & 111 & 89 (-19.8) & 30.6 & 126 & 95 (-24.6) & 18.1 \\
  tof\_5 & 9 & 84 & 65 (-22.6) & 2.8 & 99 & 78 (-21.2) & 1.7 & 105 & 84 (-20.0) & 1.7 \\
  grover\_5 & 9 & 364 & 287 (-21.2) & 9.1 & 379 & 301 (-20.6) & 10.4 & 430 & 319 (-25.8) & 9.3 \\
  vbe\_adder\_3 & 10 & 105 & 81 (-22.9) & 33.2 & 111 & 85 (-23.4) & 12.3 & 126 & 93 (-26.2) & 13.8 \\
  fprenor & 10 & 229 & 179 (-21.8) & 32.8 & 250 & 188 (-24.8) & 61.0 & 274 & 191 (-30.3) & 159.0 \\
  mod\_red\_21 & 11 & 206 & 165 (-19.9) & 7.6 & 233 & 189 (-18.9) & 153.6 & 239 & 190 (-20.5) & 21.6 \\
  gf2\^4\_mult & 12 & 246 & 205 (-16.7) & 39.5 & 279 & 225 (-19.4) & 21.5 & 318 & 254 (-20.1) & 166.6 \\
  rc\_adder\_6 & 14 & 182 & 148 (-18.7) & 489.3 & 188 & 155 (-17.6) & 193.1 & 212 & 150 (-29.2) & 61.3 \\
  csla\_mux\_3 & 15 & 168 & 150 (-10.7) & 603.9 & 195 & 159 (-18.5) & 179.5 & 219 & 188 (-14.2) & 111.6 \\
  gf2\^5\_mult & 15 & 397 & 340 (-14.4) & 604.3 & 460 & 392 (-14.8) & 606.2 & 535 & 446 (-16.6) & 615.1 \\
  ham15-med & 17 & 996 & 799 (-19.8) & 602.3 & 1155 & 941 (-18.5) & 603.6 & 1275 & 1010 (-20.8) & 605.3 \\
  ham15-low & 17 & 584 & 527 (-9.8) & 614.9 & 710 & 624 (-12.1) & 611.7 & 845 & 769 (-9.0) & 612.0 \\
  gf2\^6\_mult & 18 & 599 & 511 (-14.7) & 601.7 & 695 & 599 (-13.8) & 603.5 & 773 & 630 (-18.5) & 605.4 \\
  barenco\_tof\_10 & 19 & 285 & 233 (-18.2) & 7.9 & 336 & 271 (-19.3) & 25.2 & 372 & 267 (-28.2) & 10.1 \\
  tof\_10 & 19 & 225 & 192 (-14.7) & 632.0 & 255 & 201 (-21.2) & 87.2 & 267 & 212 (-20.6) & 26.1 \\
  ham15-high & 20 & 3564 & 2987 (-16.2) & 618.3 & 4173 & 3383 (-18.9) & 609.7 & 4809 & 3901 (-18.9) & 622.6 \\
  gf2\^7\_mult & 21 & 807 & 708 (-12.3) & 611.0 & 948 & 809 (-14.7) & 610.6 & 1050 & 892 (-15.0) & 612.9 \\
  qcla\_com\_7 & 24 & 328 & 302 (-7.9) & 625.2 & 373 & 351 (-5.9) & 608.5 & 427 & 381 (-10.8) & 622.1 \\
  adder\_8 & 24 & 800 & 703 (-12.1) & 609.9 & 893 & 734 (-17.8) & 611.9 & 953 & 766 (-19.6) & 610.9 \\
  gf2\^8\_mult & 24 & 1164 & 1066 (-8.4) & 602.1 & 1338 & 1193 (-10.8) & 604.5 & 1494 & 1337 (-10.5) & 602.6 \\
  \bottomrule
  \end{tabular}
  }}
  \end{table}
  
  \begin{table}[htbp]
  \caption{Experiment 3b: Mapped Feynman cnot-depth optimization, m: cx-depth, q: qubits, t: time. }
  \label{tb:experiment3b_feynman_cxd}
  \centering
  \resizebox{\columnwidth}{!}{%
  \rotatebox{90}{
  \begin{tabular}{lllllllllll}
  \toprule
  & & \multicolumn{3}{c}{sycamore-54}& \multicolumn{3}{c}{rigetti-80}& \multicolumn{3}{c}{eagle-127}\\
  \cmidrule(lr){3-5} \cmidrule(lr){6-8} \cmidrule(lr){9-11}
  & &org & \multicolumn{2}{c}{qsynth-f}& org & \multicolumn{2}{c}{qsynth-f}& org & \multicolumn{2}{c}{qsynth-f}\\
  \cmidrule(lr){4-5} \cmidrule(lr){7-8} \cmidrule(lr){10-11}
  instance & q & m  & m (ch\%)  & t & m  & m (ch\%)  & t & m  & m (ch\%)  & t \\
  \midrule
  barenco\_tof\_3 & 5 & 43 & 37 (-14.0) & 1.0 & 43 & 35 (-18.6) & 0.8 & 50 & 36 (-28.0) & 0.7 \\
  tof\_3 & 5 & 31 & 26 (-16.1) & 0.7 & 37 & 28 (-24.3) & 0.5 & 37 & 28 (-24.3) & 0.6 \\
  qft\_4 & 5 & 97 & 67 (-30.9) & 2.3 & 100 & 73 (-27.0) & 1.7 & 104 & 79 (-24.0) & 1.7 \\
  mod5\_4 & 5 & 30 & 26 (-13.3) & 0.9 & 37 & 28 (-24.3) & 1.0 & 43 & 32 (-25.6) & 0.8 \\
  hwb6 & 7 & 204 & 149 (-27.0) & 6.6 & 215 & 172 (-20.0) & 7.5 & 238 & 185 (-22.3) & 13.5 \\
  barenco\_tof\_4 & 7 & 67 & 56 (-16.4) & 1.7 & 72 & 60 (-16.7) & 1.3 & 78 & 64 (-17.9) & 1.3 \\
  tof\_4 & 7 & 53 & 39 (-26.4) & 1.2 & 59 & 41 (-30.5) & 1.5 & 68 & 50 (-26.5) & 1.2 \\
  barenco\_tof\_5 & 9 & 97 & 78 (-19.6) & 2.8 & 110 & 88 (-20.0) & 2.2 & 128 & 91 (-28.9) & 2.1 \\
  mod\_mult\_55 & 9 & 70 & 51 (-27.1) & 2.4 & 77 & 63 (-18.2) & 3.8 & 84 & 65 (-22.6) & 4.9 \\
  tof\_5 & 9 & 75 & 53 (-29.3) & 1.5 & 94 & 71 (-24.5) & 1.4 & 99 & 77 (-22.2) & 1.5 \\
  grover\_5 & 9 & 321 & 246 (-23.4) & 7.4 & 331 & 247 (-25.4) & 8.4 & 390 & 279 (-28.5) & 7.1 \\
  vbe\_adder\_3 & 10 & 79 & 56 (-29.1) & 3.1 & 81 & 61 (-24.7) & 2.2 & 94 & 68 (-27.7) & 3.9 \\
  fprenor & 10 & 189 & 129 (-31.7) & 6.1 & 193 & 130 (-32.6) & 6.1 & 205 & 140 (-31.7) & 7.3 \\
  mod\_red\_21 & 11 & 183 & 135 (-26.2) & 4.6 & 202 & 154 (-23.8) & 12.3 & 210 & 164 (-21.9) & 6.2 \\
  gf2\^4\_mult & 12 & 207 & 155 (-25.1) & 9.1 & 195 & 155 (-20.5) & 7.4 & 208 & 165 (-20.7) & 16.8 \\
  rc\_adder\_6 & 14 & 140 & 105 (-25.0) & 7.2 & 162 & 121 (-25.3) & 6.1 & 170 & 113 (-33.5) & 9.0 \\
  csla\_mux\_3 & 15 & 103 & 92 (-10.7) & 9.5 & 127 & 101 (-20.5) & 5.6 & 137 & 128 (-6.6) & 12.0 \\
  gf2\^5\_mult & 15 & 269 & 208 (-22.7) & 14.8 & 293 & 241 (-17.7) & 66.4 & 326 & 270 (-17.2) & 603.1 \\
  ham15-med & 17 & 807 & 544 (-32.6) & 606.4 & 952 & 642 (-32.6) & 604.8 & 1015 & 651 (-35.9) & 444.0 \\
  ham15-low & 17 & 414 & 344 (-16.9) & 605.0 & 458 & 349 (-23.8) & 605.5 & 547 & 430 (-21.4) & 608.3 \\
  gf2\^6\_mult & 18 & 353 & 294 (-16.7) & 41.8 & 406 & 333 (-18.0) & 547.9 & 448 & 363 (-19.0) & 51.8 \\
  barenco\_tof\_10 & 19 & 272 & 196 (-27.9) & 6.3 & 320 & 216 (-32.5) & 9.9 & 351 & 228 (-35.0) & 7.2 \\
  tof\_10 & 19 & 196 & 157 (-19.9) & 7.5 & 229 & 162 (-29.3) & 5.9 & 251 & 194 (-22.7) & 4.6 \\
  ham15-high & 20 & 2849 & 2249 (-21.1) & 610.8 & 3322 & 2560 (-22.9) & 610.7 & 3790 & 2965 (-21.8) & 620.9 \\
  gf2\^7\_mult & 21 & 439 & 357 (-18.7) & 45.5 & 506 & 392 (-22.5) & 93.0 & 547 & 439 (-19.7) & 604.8 \\
  qcla\_com\_7 & 24 & 139 & 115 (-17.3) & 36.0 & 139 & 128 (-7.9) & 61.2 & 143 & 130 (-9.1) & 46.3 \\
  adder\_8 & 24 & 347 & 285 (-17.9) & 72.1 & 397 & 311 (-21.7) & 59.2 & 388 & 322 (-17.0) & 80.5 \\
  gf2\^8\_mult & 24 & 563 & 469 (-16.7) & 605.6 & 634 & 569 (-10.3) & 607.9 & 665 & 582 (-12.5) & 608.8 \\
  \bottomrule
  \end{tabular}
  }}
  \end{table}

\end{document}